\documentclass[hyper]{prop2015}%
\usepackage[english]{babel} 
\usepackage{bbm}

\usepackage{amsfonts,amssymb,
amsmath,amscd, mathrsfs, 
mathrsfs,delarray,subfigure}
\usepackage{xypic}

\DeclareMathOperator{\Ad}{Ad}

\DeclareMathOperator{\id}{id}

\DeclareMathOperator{\Map}{Map}

\DeclareMathOperator{\Aut}{Aut}

\DeclareMathOperator{\Diff}{Diff}

\DeclareMathOperator{\rank}{rank}

\DeclareMathOperator{\tr}{tr}

\DeclareMathOperator{\Pexp}{Pexp}
\DeclareMathOperator{\SU}{SU}
\DeclareMathOperator{\SO}{SO}

\newcommand{\varrhd}{{\,\vartriangleright\,}}

\newcommand{\varsucc}{{\,\succ\,}}

\pagespan{}{}
\category{Proceedings}
\subcategory{}
\category{Proceedings}
\keywords{Higher gauge theory, higher Chern-Simons theory, higher knots, A field theoretic route to higher knots}
\subtitle{\href{http://www.maths.dur.ac.uk/lms/109/index.html}{LMS/EPSRC Durham Symposium on Higher Structures in M-Theory}}
\title{Wilson Surfaces for Surface Knots}
\author[F. Author]{Roberto Zucchini\inst{a,}\footnote{Corresponding author e-mail:~\href{mailto:roberto.zucchini@unibo.it}{\textsf{roberto.zucchini@unibo.it}}}}
\address[1]{DIFA, University of Bologna, and INFN, Bologna, Italy}
\shortauthors{R. Zucchini}
\shortabstract
\begin{abstract}
Holonomy invariants in strict higher gauge theory have been studied
in depth, aiming to applications to higher Chern--Simons theory. 
For a flat 2--connection, the holonomy of surface knots of arbitrary
genus has been defined and its covariance properties under 1--gauge
transformation and change of base data have been determined. 
Using quandle theory, a definition of trace over a crossed module has been given that yields surface knot invariants upon application to 2--holonomies.
\end{abstract}

\begin{document}

\maketitle

\section{Introduction}

Knots are interesting in topology as well as in gauge theory \cite{Kauffman:1991ds}.

\vspace{2mm}

Ordinary knots are embeddings of $S^1$ into a 3--dimen\-sional manifold,
say $S^3$ \cite{adams2004knot,lickorish1997introduction}. Can one define higher dimensional knots 
generalizing this simple topological notion? 
In just one dimension higher there are at least two ways of doing that.

\vspace{2mm}

Since $S^1$ is the lowest dimensional non trivial sphere, one may 
define a 2--dimensional knot as an embedding $S^2$ into $S^4$.
This yields the so called 2--knots.  
Since $S^1$ is also the lowest dimensional non trivial closed oriented manifold,
one may define a 2--dimensional knot as an embedding of $S_\ell$ 
into $S^4$, where $S_\ell$ is a genus $\ell$ closed oriented surface. 
This leads to genus $\ell$ surface knots. Of course, 2--knots are just 
genus 0 surface knots. However, they have very special properties which
make a separate study meaningful. 
2- and surface knots are objects of intense investigation by topologists 
\cite{carter1998knotted,kamada2002braid}. 

\vspace{2mm}

Wilson lines \cite{Wilson:1974sk}
are relevant in the analysis of confinement in quantum chromodynamics, 
loop formulation of quantum gravity, symmetry breaking in string theory, condensed 
matter theory and knot topology. As shown in Witten's seminal work \cite{Witten:1988hf}, 
one can study knot topology in Chern-Simons theory, an instance of gauge theory, 
relying on techniques of quantum field theory.
With any knot $\xi$, one associates the Wilson line
\begin{equation}
W_R(\xi)=\tr_R\!\bigg[\Pexp\bigg(-\int_\xi A\bigg)\bigg].
\end{equation}
where $R$ is a representation of the gauge group $G$. Chern--Simons correlators of Wilson line 
operators provide classic knot invariants. 


\vspace{2mm}

Wilson surfaces \cite{Chepelev:2001mg,Alekseev:2015hda}
may turn out to be relevant in the study of non perturbative aspects of higher 
gauge theory, brane theory, quantum gravity and higher knot topology. Following Witten's paradigm, 
one can presumably study 2-- or surface knot topology computing correlators of knot Wilson surfaces 
in an appropriate higher version of Chern--Simons theory, an instance of higher gauge 
theory \cite{Baez:2002jn,Baez:2010ya}, 
using again  techniques of quantum field theory. To this end, one needs to associate 
with any surface knot $\varXi$ a Wilson surface
\begin{equation}
W(\varXi)=?~,
\end{equation}
whose expressions is at this point to be found. 
In this communication, we shall present a proposal for a definition of the Wilson surfaces 
$W(\varXi)$ in higher gauge theory based mainly on our work 
\cite{Zucchini:2015wba,Zucchini:2015xba}. 

\vspace{2mm}

\noindent
The problem has two parts:
\begin{enumerate}[i)]

\vspace{2mm}

\item {\it define surface knot holonomy};

\vspace{2mm}

\item {\it define higher invariant traces}.

\vspace{2mm}

\end{enumerate}
Parallel transport and holonomy are related but distinguished,
holonomy being a special case of parallel transport. 

\vspace{2mm}

\noindent
Earlier endeavours on  higher parallel transport includes the work of Caetano and Picken 
\cite{Caetano:1993zf}, Baez and Schreiber \cite{Baez:2004in,Baez:2005qu}
Schreiber and Waldorf \cite{Schreiber:0705.0452,Schreiber:0802.0663,Schreiber:2008aa}, 
Faria Martins and Picken \cite{Martins:2007uki,Martins:2011:3309}, 
Chatterjee, Lahiri and Sengupta \cite{Chatterjee:2009ne,Chatterjee:2014pna,Chatterjee:2010xa} 
Soncini and Zucchini \cite{Soncini:2014zra}, Abbaspour and Wagemann \cite{Abbaspour:1202.2292}
and  Arias Abad and Schaetz \cite{Abad:1404.0729,Abad:1404.0727}.  
Earlier results on higher holonomy were obtained by Cattaneo and Rossi \cite{Cattaneo:2002tk}
and Faria Martins and Picken \cite{Martins:2007uki,Martins:2011:3309}. 

\vspace{2mm}

Following \cite{Zucchini:2015wba,Zucchini:2015xba}, we 
shall present a framework for the construction of holonomy invariants
of knots and surface knots. In a nutshell, our
strategy rests on describing knots by parametrized curves and surface knots by 
parametrized surfaces.
We outline it below, assuming that the reader is familiar with the basic ideas 
of strict higher gauge theory. In any case, those notion will be reviewed in greater 
detail in subsequent sections. 

\vspace{2mm}

In a manifold $M$, a curve $\gamma:p_0\rightarrow p_1$ is a parame\-trized path joining two points.
A homotopy $h:\gamma_0\Rightarrow \gamma_1$ of two curves is a parame\-trized path joining 
those curves. Curves can be composed by concatenation and inverted. The resulting operations 
make curves modulo homotopy a groupoid, the fundamental 1-groupoid $(M,P^0{}_1M)$ of $M$.

\vspace{2mm}

In ordinary gauge theory with gauge Lie group $G$, given a flat gauge field $A$ 
one can construct a gauge covariant and homotopy invariant parallel transport 
functor 
\begin{align}
&F_A:(M,P^0{}_1M)\rightarrow BG, 
\\
&\gamma\rightarrow F_A(\gamma),\notag
\end{align}
where $BG$ is the delooping of $G$, that is $G$ seen as the morphism group of a 
one--object groupoid. 

\vspace{2mm}

With a knot $\xi$ based at $p$ defined up to ambient isotopy one can associate a curve 
$\gamma_\xi:p\rightarrow p$ defined up to homotopy and with this the holonomy 
\begin{equation}
F_A(\xi)=F_A(\gamma_\xi).
\end{equation}
One can check $F_A(\xi)$ is base point and isotopy invariant and gauge independent
up to conjugation. 
Using invariant traces, one can extract an invariant from the holonomy $F_A(\xi)$.

\vspace{2mm}

A `gentle' generalization
of the above construction for surface knots is the following.

\vspace{2mm}

A curve $\gamma:p_0\rightarrow p_1$ is a parametrized path joining two points. 
A surface $\varSigma:\gamma_0\Rightarrow \gamma_1$ is a parametrized path joining two curves 
in a manifold $M$. A thin homotopy $h:\gamma_0\Rightarrow \gamma_1$ of two curves
is a parametrized path joining those curves with degenerate (less than two--dimensional) range  
A homotopy $H:\varSigma_0\Rrightarrow \varSigma_1$ of two surfaces is a parametrized path joining  
those surfaces.Curves can be composed by concatenation and inverted. 
Surfaces can be composed by concatenation and inverted in two distinct ways, usually called 
horizontal and vertical. The resulting operations 
make curves modulo thin homotopy and surfaces modulo homotopy a 2--groupoid, 
fundamental 2--groupoid $(M,P_1M,P^0{}_2M)$.

\vspace{2mm}

In strict higher gauge theory with gauge Lie crossed module $(G,H)$, given a flat higher gauge 
field pair $A,B$ one can construct a gauge covariant and (thin) homotopy invariant parallel transport 
2--functor 
\begin{align}
&F_{A,B}:(M,P_1M,P^0{}_2M)\rightarrow B(G,H), 
\\
&\gamma\rightarrow F_A(\gamma), \quad \varSigma\rightarrow F_{A,B}(\varSigma).
\nonumber
\end{align}

\vspace{2mm}

With a knot $\xi$ based at $p$ and a surface knot $\varXi$ based at a genus dependent fundamental 
polygon $\tau$ stemming from cutting the image of $\varXi$ along standard $a$ and $b$ cycles, both 
defined up to ambient isotopy, one can associate a curve $\gamma_\xi:p\rightarrow p$ and a
surface $\varSigma_\varXi:\iota_p\Rightarrow \tau$ up to (thin) homotopy and from this the 
holonomy
\begin{equation}
F_A(\xi)=F_A(\gamma_\xi), \quad F_{A,B}(\varXi)=F_{A,B}(\varSigma_\varXi).
\end{equation}
One can check that $F_A(\xi)$ is base data and isotopy invariant and gauge independent
up to the appropriate form of crossed module conjugation. Using higher invariant traces, one 
can extract invariants from the holonomy $F_A(\xi)$ and $F_{A,B}(\varXi)$. 

There are open issues to be solved. It can be shown that surface knot holonomy necessarily lies 
in the kernel of the target map $H\longrightarrow\!\!\!\!\!\!\!\!\!\!\!{}^t\,\,\,\,\,\,\,G$ 
of the Lie crossed module $(G,H)$ and so is central. Thus, for many Lie crossed module 
this holonomy may turn out to be trivial. 
The existence of non trivial higher traces on $(G,H)$ is also to be ascertained. This is an 
problem that can be formalized using higher quandle theory 
(see Crans \cite{Crans:2004ve} and Crans and Wagemann\cite{Crans:1310.4705}). 

From a quantum field theoretic point of view, the most delicate question remains obtaining 
surface knot invariants from a 4--dimensional higher Chern--Simons theory as proposed 
by Zucchini \cite{Zucchini:2011aa,Zucchini:2015ohw} and Soncini and Zucchini 
\cite{Soncini:2014ara}. There are problems with 
the definition of Wilson surface insertions in the quantum theory, which we shall point out in due 
course.

\section{Curves, surfaces and homotopy} \label{sec:holo}

Closed curves and surfaces describe knots and surface knots
in an ambient manifold $M$. 

Curves and surfaces are smoothly parametrized subsets of $M$. 
They can be composed and inverted in various ways. In order to preserve 
smoothness, it is sufficient to require that their parametrization
has sitting instants. 
A smooth map $f:S\times\mathbbm{R}\rightarrow T$, where $S$ and $T$ are manifolds,
has sitting instants if
\begin{align}
&f(-,x)=f(-,0) \quad\text{for $x<\epsilon$}, \\
&f(-,x)=f(-,1) \quad \text{for $x>1-\epsilon$}, \nonumber
\end{align}
for some number $\epsilon$ such that 
$0<\epsilon<1/2$. In what follows, all maps will be tacitly assumed 
to have sitting instants for each factor $\mathbbm{R}$ of their domains. 

\vspace{2mm}

Formally, curves and surfaces are defined as follows. 

\vspace{2mm}

\noindent
{\it For any two points $p_0,p_1\in M$, a curve $\gamma:p_0\rightarrow p_1$ in $M$ is 
a map $\gamma:\mathbbm{R}\rightarrow M$ such that}
\begin{align}
\gamma(0)=p_0,\quad \gamma(1)=p_1.
\end{align}

\vspace{2mm}

\noindent
{\it 
For any two points $p_0,p_1\in M$ and any two curves $\gamma_0,\gamma_1:p_0\rightarrow p_1$ of $M$, a surface  
$\varSigma:\gamma_0\Rightarrow \gamma_1$ of $M$ is a map $\varSigma:$ $\mathbbm{R}^2\rightarrow M$  such that}
\begin{align}
&\varSigma(0,y)=p_0,\quad \varSigma(1,y)=p_1,
\\
&\varSigma(x,0)=\gamma_0(x),\quad \varSigma(x,1)=\gamma_1(x)   \nonumber
\end{align}

\vspace{2mm}

\noindent
Curve and surfaces can be combined through a set of natural operations
based on the intuitive idea of concatenation. We begin by introducing the basic 
operations with curves. 

\vspace{2mm}

\noindent
{\it For a point $p$, the unit curve of $p$ is the curve $\iota_p:p\rightarrow p$ defined by 
\begin{align}
\iota_p(x)=p.
\end{align}
For a curve $\gamma:p_0\rightarrow p_1$, the inverse curve of $\gamma$ is 
the curve $\gamma^{-1_\circ}:p_1\rightarrow p_0$ given by 
\begin{align}
\gamma^{-1_\circ}(x)=\gamma(1-x).
\end{align}
For two curves $\gamma_1:p_0\rightarrow p_1$, $\gamma_2:p_1\rightarrow p_2$, the composition 
of $\gamma_1$, $\gamma_2$ is the curve $\gamma_2\circ\gamma_1:p_0\rightarrow p_2$ piecewise given by}
\begin{align}
&\gamma_2\circ\gamma_1(x)=\gamma_1(2x) \qquad \text{for $x\leq 1/2$},
\\
&\gamma_2\circ\gamma_1(x)=\gamma_2(2x-1) \qquad \text{for $x\geq 1/2$}.\notag
\vphantom{\Big]}
\end{align}

\vspace{2mm}

\noindent 
We introduce next the basic operations with surfaces. These turn out to be of two types,
called horizontal and vertical.

\vspace{2mm}

\noindent
{\it For a curve $\gamma:p_0\rightarrow p_1$, the unit surface of $\gamma$ is the surface
$I_\gamma:\gamma\Rightarrow\gamma$ defined by 
\begin{align}
I_\gamma(x,y)=\gamma(x).
\end{align}
For a surface $\varSigma:\gamma_0\Rightarrow\gamma_1$, the vertical inverse 
of $\varSigma$ is the surface $\varSigma^{-1_\bullet}:\gamma_1\Rightarrow\gamma_0$ defined by 
\begin{align}
\varSigma^{-1_\bullet}(x,y)=\varSigma(x,1-y).
\end{align}
For two surfaces $\varSigma_1:\gamma_0\Rightarrow\gamma_1$, $\varSigma_2:\gamma_1\Rightarrow\gamma_2$, 
the vertical composition of $\varSigma_1$, $\varSigma_2$ is the surface
$\varSigma_2\bullet\varSigma_1:\gamma_0\Rightarrow \gamma_2$ given by 
\begin{align}
&\varSigma_2\bullet\varSigma_1(x,y)=\varSigma_1(x,2y) \qquad \text{for $y\leq 1/2$},
\vphantom{\Big]}
\\
&\varSigma_2\bullet\varSigma_1(x,y)=\varSigma_2(x,2y-1) \qquad \text{for $y\geq 1/2$}.\notag
\vphantom{\Big]}
\end{align}
For a surface $\varSigma:\gamma_0\Rightarrow\gamma_1$, the horizontal inverse of $\varSigma$
is the surface $\varSigma^{-1_\circ}:\gamma_0{}^{-1_\circ}\Rightarrow \gamma_1{}^{-1_\circ}$ defined by 
\begin{align}
\varSigma^{-1_\circ}(x,y)=\varSigma(1-x,y).
\end{align}
For two surfaces $\varSigma_1:\gamma_0\Rightarrow\gamma_1$, $\varSigma_2:\gamma_2\Rightarrow\gamma_3$, 
the horizontal composition of $\varSigma_1$, $\varSigma_2$ is the surface
$\varSigma_2\circ\varSigma_1:\gamma_2\circ\gamma_0\Rightarrow \gamma_3\circ\gamma_1$ given by}
\begin{align}
&\varSigma_2\circ\varSigma_1(x,y)=\varSigma_1(2x,y) \qquad \text{for $x\leq 1/2$},
\vphantom{\Big]}
\\
&\varSigma_2\circ\varSigma_1(x,y)=\varSigma_2(2x-1,y) \qquad \text{for $x\geq 1/2$}.\notag
\vphantom{\Big]}
\end{align}

\vspace{2mm}

Unfortunately, these operations are not nice enough; 
associativity and invertibility fail to hold in general. The operation are 
in fact nice only up to homotopy.

\vspace{2mm}

\noindent
{\it A homotopy $h:\gamma_0\Rightarrow \gamma_1$ 
of two curves $\gamma_0,\gamma_1:p_0\rightarrow p_1$ of $M$ with the same 
end-points is a map $h:\mathbbm{R}^2\rightarrow M$ of $M$ such that 
\begin{align}
&h(0,y)=p_0,\quad h(1,y)=p_1,
\\
&h(x,0)=\gamma_0(x),\quad h(x,1)=\gamma_1(x).
\nonumber
\end{align}
The homotopy is thin if in addition $\rank dh(x,y)<2$. (Thin) homotopy of curves 
is an equivalence relation.}

\vspace{2mm}

\noindent
{\it A homotopy $H:\varSigma_0\Rrightarrow\varSigma_1$ of two surfaces 
$\varSigma_0:\gamma_0\Rightarrow\gamma_1$, $\varSigma_1:\gamma_2\Rightarrow\gamma_3$, 
where $\gamma_0,\gamma_1,\gamma_2,\gamma_3:p_0\rightarrow p_1$ are four curves with the 
same end-points, is a map $H:\mathbbm{R}^3\rightarrow M$ such that $\rank dH(x,0,z)$, $\rank dH(x,1,z)\leq 1$ and 
\begin{align}
&H(0,y,z)=p_0,\quad H(1,y,z)=p_1, 
\\
&H(x,y,0)=\varSigma_0(x,y), \quad H(x,y,1)=\varSigma_1(x,y).
\nonumber
\end{align}
The homotopy is thin if $\rank dH(x,y,z)<3$. (Thin) homotopy of surfaces is an equivalence relation.}

\vspace{2mm}

\noindent
Let us denote by $\Pi_1M$, $\Pi_2M$ the sets of all curves and surfaces of $M$,
by $P_1M$ and $P^0{}_1M$ the sets of thin homotopy and homotopy classes of curves
and $P_2M$ and $P^0{}_2M$ the sets of thin homotopy and homotopy classes of surfaces,
respectively. The following results are basic. 

\vspace{2mm}

\noindent
{\it $(M,P_1M)$ and $(M,P^0{}_1M)$ with the operations induced by those of $\Pi_1M$
are groupoids, the path and fundamental groupoids of $M$.}

\vspace{2mm}

\noindent
{\it $(M,P_1M,P_2M)$ and $(M,P_1M,P^0{}_2M)$ with the operations induced by those of 
$\Pi_1M$ and $\Pi_2M$ are 2--groupoids, the path and fundamental 2--groupoids of $M$.}

\section{Higher parallel transport} \label{sec:partr}

In gauge theory, holonomy is a special case of parallel transport. Therefore, it is necessary 
to examine in some detail the definition and the properties of the latter.
We begin be reviewing parallel transport in ordinary gauge theory and then 
we introduce and describe parallel transport in higher gauge theory. 

Let $G$ be a Lie group with Lie algebra $\mathfrak{g}$ and let $M$ be a manifold.
Consider an ordinary gauge theory on the trivial principal $G$--bundle $M\times G$.

\vspace{2mm}

\noindent
{\it A $G$--connection on $M$ is a  $\mathfrak{g}$--valued 
$1$--form $\theta\in\Omega^1(M,\mathfrak{g})$. $\theta$ is flat if}
\begin{align}
{\rm d}\theta+\frac{1}{2}[\theta,\theta]=0. 
\label{twoholo7}
\end{align}

\vspace{2mm}

\noindent
Parallel transport requires a $G$--connection on $M$ as input datum. 

\vspace{2mm}

\noindent
{\it For a curve $\gamma$ of $M$, the parallel transport along $\gamma$}
is the element $F_\theta(\gamma)\in G$ defined by
\begin{align}
F_\theta(\gamma)=u(1),
\label{twofholo1}
\end{align}
{\it where $u:\mathbbm{R}\rightarrow G$ is the unique solution of the differential problem}
\begin{align}
{\rm d}_xu(x)u(x)^{-1}=-\gamma^*\theta_x(x), \quad u(0)=1_G.
\label{twofholo2}
\end{align}

\vspace{2mm}

\noindent
The first relevant property of parallel transport is its consistency with
the operations with curves defined in Section \ref{sec:holo}. 

\vspace{2mm}

\noindent
{\it For any point $p$ and any curves $\gamma,\gamma_1,\gamma_2$, one has}
\begin{align}
\label{ptc1}
&F_\theta(\iota_p)=1_G,
\\
\label{ptc2}
&F_\theta(\gamma^{-1_\circ})=F_\theta(\gamma)^{-1},
\\
\label{ptc3}
&F_\theta(\gamma_2\circ\gamma_1)=F_\theta(\gamma_2)F_\theta(\gamma_1).
\end{align}
{\it whenever defined.}

\vspace{2mm}

\noindent
The second relevant property of parallel transport is its compatibility 
with homotopy of curves as defined in Section \ref{sec:holo}. 

\vspace{2mm}

\noindent
{\it For any two thinly homotopic curves $\gamma_0$, $\gamma_1$, one has 
\begin{align}
&F_\theta(\gamma_1)=F_\theta(\gamma_0). 
\label{twoholo6}
\end{align}
When $\theta$ is flat, \eqref{twoholo6} holds also when 
$\gamma_0$, $\gamma_1$ are homotopic. }

\vspace{2mm}

\noindent
Parallel transport has an elegant categorical interpretation. 

\vspace{2mm}

\noindent
{\it Parallel transport yields a functor $\bar F_\theta:(M,P_1M)\rightarrow BG$ 
from the path groupoid $(M,P_1M)$ of $M$ into $BG$. 
For flat $\theta$, parallel transport yields a functor $\bar F^0{}_\theta:(M,P^0{}_1M)\rightarrow BG$ 
from the fundamental groupoid $(M,P^0{}_1M)$ of $M$ into $BG$.}

\vspace{2mm}

Any meaningful gauge theoretic construction should be gauge covariant in the appropriate 
sense. Parallel transport has also this property. 

\vspace{2mm}

\noindent
{\it A $G$--gauge transformation is just 
a $G$--valued mapping $g\in\Map(M,G)$.}

\vspace{2mm}

\noindent
Gauge transformations act on connections in the well--known manner. 

\vspace{2mm}

\noindent
{\it The gauge transform of the $G$--connection $\theta$ is the $G$--connection}
\begin{align}
&{}^g\theta=\Ad g (\theta)-{\rm d}gg^{-1}.
\label{gauholo1}
\end{align}
{\it If $\theta$ is flat, ${}^g\theta$ is flat, too.}

\vspace{2mm}

\noindent
Parallel transport has simple covariance properties under gauge transformation.

\vspace{2mm}

\noindent
{\it For any curve $\gamma:p_0\rightarrow p_1$ of $M$, }
\begin{align}
F_{{}^g\theta}(\gamma)=g(p_1)F_\theta(\gamma)g(p_0)^{-1}.
\vphantom{\Big]}
\label{gauholo2}
\end{align}

\vspace{2mm}

\noindent
Gauge transformation of parallel transport 
also has an elegant categorical interpretation. 

\vspace{2mm}

\noindent
{\it A gauge transformation $g$ encodes a natural transformation 
$\bar F_\theta\Rightarrow \bar F_{{}^g\theta}$ of parallel transport functors.
When $\theta$ is flat, $g$ encodes a natural transformation 
$\bar F^0{}_\theta\Rightarrow \bar F^0{}_{{}^g\theta}$ of flat parallel transport functors.}

\vspace{2mm}

An appropriate form of parallel transport
can be defined also in strict higher gauge theory. The intuitive idea
of the construction is still simple, though the technical details are much more 
involved.

\vspace{2mm}

\noindent
Let $K$ be a strict Lie $2$ group with strict Lie $2$--algebra $\mathfrak{k}$ and let $M$ be a manifold.
Consider a higher gauge theory on the trivial principal $K$--$2$--bundle $M\times K$.
As it is natural and convenient, we shall view the Lie $2$--group $K$ as a Lie crossed module 
$H\longrightarrow\!\!\!\!\!\!\!\!\!\!\!{}^t\,\,\,\,\,\,\,G
\longrightarrow\!\!\!\!\!\!\!\!\!\!\!\!\!{}^m\,\,\,\,\Aut(H)$ and the Lie $2$--algebra $\mathfrak{k}$ 
as the differential Lie crossed module 
$\mathfrak{h}\longrightarrow\!\!\!\!\!\!\!\!\!\!\!{}^{\dot t}\,\,\,\,\,\,\,\mathfrak{g}
\longrightarrow\!\!\!\!\!\!\!\!\!\!\!\!\!\!{}^{\widehat m}\,\,\,\,\,\,\mathfrak{der}(\mathfrak{h})$
corresponding to it.

\vspace{2mm}

\noindent
{\it A $(G,H)$--$2$--connection on $M$ is a pair formed by a $\mathfrak{g}$--valued $1$--form
$\theta\in\Omega^1(M,\mathfrak{g})$ and a $\mathfrak{h}$--valued $2$--form 
$\varUpsilon\in\Omega^2(M,\mathfrak{h})$ such that}
\begin{align}
{\rm d}\theta+\frac{1}{2}[\theta,\theta]-\dot t(\varUpsilon)=0. 
\label{twoholo8}
\end{align}
{\it (vanishing fake curvature condition). $(\theta,\varUpsilon)$ is flat if}
\begin{align}
{\rm d}\varUpsilon+\widehat{m}(\theta,\varUpsilon)=0.
\label{twoholo27}
\end{align}

\vspace{2mm}

\noindent
Analogously to the ordinary case, 
higher parallel transport requires a $(G,H)$ connection on $M$ as input datum. 

\vspace{2mm}

\noindent
{\it For a curve $\gamma$ of $M$, the parallel transport $F_\theta(\gamma)\in G$ 
is constructed as done earlier for the $G$--connection $\theta$.
For a surface $\varSigma$ of $M$, the parallel transport along $\varSigma$
is the element $F_{\theta,\varUpsilon}(\varSigma)\in H$ defined by }
\begin{align}
F_{\theta,\varUpsilon}(\varSigma)=E(0,1), 
\label{twofholo3}
\end{align}
{\it where $E:\mathbbm{R}^2\rightarrow H$ is the unique solution of the two step 
differential problem}
\begin{align}
&\partial_xu(x,y)u(x,y)^{-1}=-\varSigma^*\theta_x(x,y),\quad u(1,y)=1_G,
\label{cycle36}
\\
&\partial_yv(x,y)v(x,y)^{-1}=-\varSigma^*\theta_y(x,y),\,\,\quad v(x,0)=1_G,
\label{cycle37}
\\
&\partial_x(\partial_yE(x,y)E(x,y)^{-1})=
\\
&\qquad\qquad=-\dot m(v(1,y)^{-1}u(x,y)^{-1})(\varSigma^*\varUpsilon_{xy}(x,y)) 
 ~~\text{or}
\nonumber 
\\
&\partial_y(E(x,y)^{-1}\partial_xE(x,y))=
\nonumber
\\
&\qquad\qquad=-\dot m(u(x,0)^{-1}v(x,y)^{-1})(\varSigma^*\varUpsilon_{xy}(x,y)),
\nonumber
\\
&\hspace{5cm} E(1,y)=E(x,0)=1_H
\nonumber
\end{align}
{\it with $u,v:\mathbbm{R}^2\rightarrow G$.}

\vspace{2mm}

\noindent
The two forms of the differential problem for $E$ are equivalent: 
any solution of one is automatically a solution of the other.

\vspace{2mm}

\noindent
Higher parallel transport has several remarkable properties which
extend those of the ordinary case. 
First, higher parallel transport along surfaces is compatible with that
along their end-curves.

\vspace{2mm}

\noindent
{\it For a surface 
$\varSigma:\gamma_0\Rightarrow \gamma_1$ joining the curve $\gamma_0$ to the curve $\gamma_1$, }
\begin{align}
&F_\theta(\gamma_1)=t(F_{\theta,\varUpsilon}(\varSigma))F_\theta(\gamma_0).
\label{twoholo14}
\end{align}

\vspace{2mm}

\noindent
Second, higher parallel transport is consistent with
the operations with curves and surfaces defined in Section \ref{sec:holo}. 

\vspace{2mm}

\noindent
{\it For any point $p$, any curves $\gamma,\gamma_1,\gamma_2$ and any surfaces 
$\varSigma,\varSigma_1,\varSigma_2$, relations \eqref{ptc1}--\eqref{ptc3}
and the further relations}
\begin{align}
&F_{\theta,\varUpsilon}(I_\gamma)=1_H,
\\
&F_{\theta,\varUpsilon}(\varSigma^{-1\bullet})=F_{\theta,\varUpsilon}(\varSigma)^{-1}, 
\\
&F_{\theta,\varUpsilon}(\varSigma_2\bullet\varSigma_1)
=F_{\theta,\varUpsilon}(\varSigma_2)F_{\theta,\varUpsilon}(\varSigma_1),
\\
&F_{\theta,\varUpsilon}(\varSigma^{-1\circ})
=m(F_\theta(\gamma_0)^{-1})(F_{\theta,\varUpsilon}(\varSigma)^{-1}),
\\
&F_{\theta,\varUpsilon}(\varSigma_2\circ\varSigma_1)
=F_{\theta,\varUpsilon}(\varSigma_2)m(F_\theta(\gamma_2))(F_{\theta,\varUpsilon}(\varSigma_1)),
\end{align}
{\it hold whenever defined, where in the last two identities $\varSigma:\gamma_0\Rightarrow\gamma_1$
and $\varSigma_2:\gamma_2\Rightarrow\gamma_3$.}

\vspace{2mm}

\noindent
Third, higher parallel transport is compatible
with homotopy of curves and surfaces, as defined again in Section \ref{sec:holo},
in the following sense. 

\vspace{2mm}

\noindent
{\it For any two thinly homotopic curves $\gamma_0$, $\gamma_1$}
\begin{align}
\label{twoholo6/2}
&F_\theta(\gamma_1)=F_\theta(\gamma_0). 
\end{align}
{\it For any two thinly homotopic
surfaces $\varSigma_0:\gamma_{00}\Rightarrow\gamma_{01}$, $\varSigma_1:\gamma_{10}\Rightarrow\gamma_{11}$,}
\begin{align}
&F_\theta(\gamma_{10})=F_\theta(\gamma_{00}),
\\
&F_\theta(\gamma_{11})=F_\theta(\gamma_{01}),
\\
&F_{\theta,\varUpsilon}(\varSigma_1)=F_{\theta,\varUpsilon}(\varSigma_0).
\end{align}
{\it The same relations hold if $(\theta,\varUpsilon)$ is flat and $\varSigma_0$, $\varSigma_1$ are homotopic.}

\vspace{2mm}

\noindent
Higher parallel transport has an elegant $2$--categorical interpretation. 

\vspace{2mm}

\noindent
{\it Higher parallel transport is equivalent to  a strict $2$--fun\-ctor 
$\bar F_{\theta,\varUpsilon}:(M,P_1M,P_2M)\rightarrow B(G,H)$ 
from the path $2$--groupoid  $(M,P_1M,P_2M)$ of $M$ into $B(G,H)$.
For a flat $(\theta,\varUpsilon)$, higher parallel transport is likewise equivalent to a 
strict $2$--func\-tor $\bar F^0{}_{\theta,\varUpsilon}:(M,P_1M,P^0{}_2M)\rightarrow B(G,H)$ 
from the fundamental $2$--groupoid $(M,P_1M,P^0{}_2M)$ of $M$ into $B(G,H)$.}

\vspace{2mm}

\noindent
Here, with an abuse of notation, $B(G,H)$ stands for the 
delooping of the strict Lie 2--group corresponding to the Lie crossed module $(G,H)$. 

\vspace{2mm}

Analogously to ordinary gauge theory, higher parallel transport
is gauge covariant in the appropriate higher sense. 

\vspace{2mm}

\noindent
{\it A $(G,H)$--$1$--gauge transformation is a pair of a $G$--valued map $g\in\Map(M,G)$
and an $\mathfrak{h}$--valued $1$--form $J\in\Omega^1(M,\mathfrak{h})$.}

\vspace{2mm}

\noindent
$1$--gauge transformations act on $2$ connections.

\vspace{2mm}

\noindent
{\it The 1--gauge transform of the $(G,H)$--$2$--connection $(\theta,\varUpsilon)$
is the $2$--connection}
\begin{align}
&{}^{g,J}\theta=\Ad g(\theta)-dgg^{-1}-\dot t(J), 
\vphantom{\Big]}
\label{gauholo3}
\\
&{}^{g,J}\varUpsilon=\dot m(g)(\varUpsilon)-{\rm d}J-\frac{1}{2}[J,J]\,-
\label{gauholo4}
\\
&\hspace{3cm}-\widehat{m}(\Ad g(\theta) -{\rm d}gg^{-1}-\dot t(J),J).
\vphantom{\Big]}
\nonumber 
\end{align}
{\it If $(\theta,\varUpsilon)$ is flat, $({}^{g,J}\theta,{}^{g,J}\varUpsilon)$ is flat, too.}

\vspace{2mm}

\noindent
There exits a notion of parallel transport for $1$--gauge transformations
similarly to $2$--connections. 

\vspace{2mm}

\noindent
{\it For a curve $\gamma$ of $M$, the gauge parallel transport along $\gamma$
is the element $G_{g,J;\theta}(\gamma)\in H$ given by} 
\begin{align}
G_{g,J;\theta}(\gamma)=\varLambda(0),
\end{align}
{\it where $\varLambda:\mathbbm{R}\rightarrow H$ is the unique solution of the two--step 
differential problem}
\begin{align}
&{\rm d}_xu(x)u(x)^{-1}=-\gamma^*\theta_x(x), \quad u(1)=1_G,
\vphantom{\Big]}
\\
&\varLambda(x)^{-1}{\rm d}_x\varLambda(x)=-\dot m(u(x)^{-1}\gamma^*g(x)^{-1})(\gamma^*J_x(x)),
\\
&\hspace{6cm}\varLambda(1)=1_H.
\nonumber
\vphantom{\Big]}
\nonumber 
\end{align}

\vspace{2mm}

\noindent
As ordinary parallel transport, gauge parallel transport is consistent with
the operations with curves defined in Section \ref{sec:holo}. 

\vspace{2mm}

\noindent
{\it For any point $p$ and any curves $\gamma,\gamma_1,\gamma_2$, one has}
\begin{align}
&G_{g,J;\theta}(\iota_p)=1_H,
\\
&G_{g,J;\theta}(\gamma^{-1_\circ})=m(F_\theta(\gamma)^{-1})(G_{g,J;\theta}(\gamma)^{-1}),
\\
&G_{g,J;\theta}(\gamma_2\circ\gamma_1)
=G_{g,J;\theta}(\gamma_2)m(F_\theta(\gamma_2))(G_{g,J;\theta}(\gamma_1)).
\end{align}
{\it  whenever defined.}

\vspace{2mm}

\noindent
Again as ordinary parallel transport, gauge parallel transport is compatible 
with homotopy of curves as defined in Section \ref{sec:holo}. 

\vspace{2mm}

\noindent
{\it For any two thinly homotopic curves $\gamma_0$, $\gamma_1$, one has }
\begin{align}
G_{g,J;\theta}(\gamma_1)=G_{g,J;\theta}(\gamma_0). 
\label{gauholo6/2}
\end{align}

\vspace{2mm}

\noindent
The reason why we introduced gauge parallel transport is that it enters in 
the $1$--gauge covariance relation of higher parallel transport in a non trivial manner. 

\vspace{2mm}

\noindent
{\it For any curve $\gamma:p_0\rightarrow p_1$, one has}
\begin{align}
\label{gauholo16/1}
F_{{}^{g,J}\theta}(\gamma)
=g(p_1)t(G_{g,J;\theta}(\gamma))F_\theta(\gamma)g(p_0)^{-1}.
\end{align}
{\it For any two curves $\gamma_0,\gamma_1:p_0\rightarrow p_1$ and any surface
$\varSigma:\gamma_0\Rightarrow\gamma_1$, one has}
\begin{align}
\label{gauholo7}
&F_{{}^{g,J}\theta,{}^{g,J}\varUpsilon}(\varSigma)=
\\
&\hspace{1cm}=m(g(p_1))\big(G_{g,J;\theta}(\gamma_1)
F_{\theta,\varUpsilon}(\varSigma)G_{g,J;\theta}(\gamma_0)^{-1}\big).
\nonumber
\end{align}

\vspace{2mm}

\noindent
Gauge parallel transport has as expected a categorical interpretation. 

\vspace{2mm}

\noindent
{\it Gauge parallel transport defines a pseudonatural transformation 
$\bar G_{g,J;\theta}:\bar F_{\theta,\varUpsilon}\Rightarrow \bar F_{{}^{g,J}\theta,{}^{g,J}\varUpsilon}$
of parallel transport $2$--functors. 
If $(\theta,\varUpsilon)$ is a flat $2$--connection, 
gauge parallel transport defines a pseudonatural transformation 
$\bar G^0{}_{g,J;\theta}:\bar F^0{}_{\theta,\varUpsilon}\Rightarrow \bar F^0{}_{{}^{g,J}\theta,{}^{g,J}\varUpsilon}$
of flat parallel transport $2$--functors.}

\vspace{2mm}

\noindent
Higher gauge theory is characterized also by gauge for gauge symmetry.  

\vspace{2mm}

\noindent
{\it A $(G,H)$--$2$--gauge transformation is just a mapping $\varOmega\in\Map(M,H)$.}

\vspace{2mm}

\noindent
$(G,H)$--$2$--gauge transformations describe gauge transformations of $(G,H)$--$1$-- gauge transformations
depending on an assigned $(G,H)$--$2$--connection $(\theta,\varUpsilon)$. 
They encode modifications 
$\bar G_{g,J;\theta}\Rrightarrow\bar G_{{}^{\tilde \varOmega} g_{|\theta},{}^{\tilde \varOmega}J_{|\theta};\theta}$ 
of gauge pseudonatural transformations of parallel transport functors.
The apparently have no role in knot holonomy.

\section{$C$-- and $S$--knots} \label{sec:csknots}

Knots are embeddings of a fixed closed model manifold into an ambient manifold $M$.
Thus, knots are not simply subsets of $M$ but mappings into $M$.
Knots differing by an ambient isotopy are identified. 

\vspace{2mm}

The simplest closed model manifold is the oriented circle $C$. 

\vspace{2mm}

\noindent
{\it A $C$--marking of $C$ is a pointing $p_C\in C$ of $C$.}

\vspace{2mm}

\noindent
{\it A $C$--marking of an oriented manifold $M$ is a pointing $p_M\in M$ of $M$.} 

\vspace{2mm}

\noindent
$C$--knots are circles embedded in $M$. 

\vspace{2mm}

\noindent
{\it A marked $C$--knot of $M$ is embedding $\xi:C\rightarrow M$
of the circle $C$ into $M$ such that}
\begin{align}
&\xi(p_C)=p_M.
\end{align}

\vspace{2mm}

\noindent
Ambient isotopy is the natural notion of mutual deformability of marked $C$--knots. 

\vspace{2mm}

\noindent
{\it Two marked $C$--knots $\xi_0,\xi_1$ are ambient isotopic if
there is a smooth family $F_z\in\Diff_+(M)$, $z\in\mathbbm{R}$, of orientation preserving 
diffeomorphisms such that}
\begin{align}
&F_0=\id_M,\quad \xi_1=F_1\circ \xi_0, 
\\
&F_z(p_M)=p_M.
\end{align}

\vspace{2mm}

In order to compute $C$--knot holonomy, we need pa\-rametrized $C$--knots.
This is achieved by assigning a curve to any marked $C$--knot as detailed next. 

\vspace{2mm}

\noindent
{\it A compatible curve in $C$ is a curve $\gamma_C:p_C\rightarrow p_C$ in $C$ such that
\begin{enumerate}[i)]

\vspace{2mm}
\item
 $I_C=\gamma_C{}^{-1}(C\setminus p_C)$ is an open interval in $\mathbbm{R}$;

\vspace{2mm}
\item
$\gamma_C|_{I_C}:I_C\rightarrow C\setminus p_C$ is an orientation preserving diffeomorphism. 

\end{enumerate}
}

\vspace{2mm}

\noindent
{\it Example}. Let $C=S^1$ be the circle standardly embedded in $\mathbbm{R}^2$ through 
\begin{equation}
s_{S^1}(\vartheta)=(\cos\vartheta,\sin\vartheta),
\end{equation}
where $\vartheta\in [0,2\pi)$, with the $C$--marking $p_{S^1}=(1,0)$. A compatible curve
$\gamma_{S^1}:\mathbbm{R}\rightarrow S^1$ is given by 
\begin{equation}
\gamma_{S^1}(x)=s_{S^1}(2\pi\alpha(x))
\label{gammas1}
\end{equation}
where $\alpha:\mathbbm{R}\rightarrow [0,1]$ is a function such that 
$d_x\alpha(x)\geq 0$ and $\alpha(x)=0$ for $x<\epsilon$ and $\alpha(x)=1$ for $x>1-\epsilon$.

\vspace{2mm}

\noindent
A curve furnishing a natural parametrization of a given marked $C$--knot 
can now be constructed. 

\vspace{2mm}

\noindent
{\it With every marked $C$--knot $\xi$ there is associated a curve  \linebreak 
$\gamma_\xi:p_M\rightarrow p_M$ given by}
\begin{align}
\gamma_\xi=\xi\circ \gamma_C.
\end{align}

\vspace{2mm}

\noindent
The curve $\gamma_\xi$ has a number of nice properties.

\vspace{2mm}

\noindent
{\it $\gamma_\xi$ is independent of the choice of the compatible curve 
$\gamma_C$ up to thin homotopy.}

\vspace{2mm}

\noindent
Note that the $C$--marking $p_C$ of $C$ is fixed here. 

\vspace{2mm}

\noindent
Ambient isotopic marked $C$--knots have homotopic cur\-ves

\vspace{2mm}

\noindent
{\it If $\xi_0$, $\xi_1$ are ambient isotopic marked $C$--knots, then $\gamma_{\xi_0}$, $\gamma_{\xi_1}$ 
are homotopic in the sense of Section \ref{sec:holo}.}

\vspace{2mm}

\noindent
It should be possible to alter marking data changing $\gamma_\xi$ at most by
a (thin) homotopy.

\vspace{2mm}

\noindent
{\it Two marked $C$--knots $\xi_0$, $\xi_1$ with respect to two distinct
$C$--marking $p_{M0}$, $p_{M1}$ of $M$ are freely ambient isotopic if there is a 
smooth family $F_z\in\Diff_+(M)$, $z\in\mathbbm{R}$, of orientation preserving 
diffeomorphisms such that}
\begin{align}
&F_0=\id_M,\quad \xi_1=F_1\circ \xi_0.
\end{align}

\vspace{2mm}

\noindent
Again, the $C$--marking $p_C$ of $C$ is fixed. 

\vspace{2mm}

\noindent
Freely ambient isotopic marked $C$--knots have homotopic curves up to conjugation.

\vspace{2mm}

\noindent
{\it If $\xi_0$, $\xi_1$ two freely ambient isotopic marked $C$--knots, there is a 
curve $\gamma_1:p_{M0}\rightarrow p_{M1}$ such that $\gamma_{\xi_0}$, 
$\gamma_1{}^{-1_\circ }\circ\gamma_{\xi_1}\circ\gamma_1$ are homotopic.}

\vspace{2mm}

\noindent
Notice that the ``compose rightmost first'' convention is used here and in the following
for curve composition. 

\vspace{2mm}

\noindent
The same embedding of $C$ into $M$ can be a marked $C$--knot in more than one way. The
corresponding curves are related in the expected manner.

\vspace{2mm}

\noindent
{\it If the embedding $\xi:C\rightarrow M$ is a marked $C$--knot with respect to two 
distinct $C$--markings $p_{C0}$, $p_{M0}$ and $p_{C1}$, $p_{M1}$ of $C$ and $M$, there exists 
a curve $\gamma_1:p_{M0}\rightarrow p_{M1}$ in $\xi(C)$ such that 
$\gamma_{\xi|0}$, $\gamma_1{}^{-1_\circ}\circ\gamma_{\xi|1}\circ\gamma_1$ are thinly homotopic.}

\vspace{2mm}

The results just expounded are standard. Our aim is finding their generalization to surface knots.
As we shall see, this task is not completely straightforward. 
Problems occur for higher genus knots. We shall propose a solution in due course. 
To this end, we need to introduce further notions. 

\vspace{2mm}

\noindent
To construct higher genus $S$--knot holonomy, it will be necessary to cut 
the model manifold $S$ along its standard $a$-- and $b$--cycles. The cuts are the images 
of spiky $C$--knots, generalized $C$--knots which are continuous but not smooth at 
the marked point. 

\vspace{2mm}

\noindent
{\it A spiky $C$--knot is an embedding $\xi:C\rightarrow M$ that obeys 
\begin{align}
\xi(p_C)=p_M
\end{align}
and is smooth on $C\setminus p_C$ with finite derivatives and non zero first derivatives 
at both ends of $C\setminus p_C$.}

\vspace{2mm}

\noindent
Note that spiky $C$--knots are marked. 

\vspace{2mm}

\noindent
With any spiky $C$--knot $\xi$, one can associate a curve 
$\gamma_\xi$ defined in the same way as above and smooth anyway.

\vspace{2mm}

\noindent
{\it For every spiky marked $C$--knot $\xi$, the curve $\gamma_\xi$ is smooth.}

\vspace{2mm}

We can now introduce $S$--knots. The next to simplest closed manifold is a 
genus $\ell_S$ closed oriented surface $S$.

\vspace{2mm}

\noindent
{\it An $S$--marking of $M$ consists of the following elements:
\begin{enumerate}[i)]

\vspace{2mm}
\item 
a $C$--marking $p_M$ of $M$;

\vspace{2mm}
\item 
a set of spiky $C$--knots $\zeta_{Mi}$ of $M$, $i=1,\dots,2\ell_S$, such that:

\vspace{2mm}
\item
 the $\zeta_{Mi}(C)$ intersect only at $p_M$;

\vspace{2mm}
\item
 there is an  embedding $\varPhi:S\rightarrow M$ with the property that
$\varPhi(p_S)=p_M$, $\varPhi\circ\zeta_{Si}=\zeta_{Mi}$. 

\end{enumerate}
}

\vspace{2mm}

\noindent
Note that the notion of $S$--marking of $S$ is compatible with that of $S$--marking of $M$
when $M=S$.

\vspace{2mm}

\noindent
$S$--knots are surfaces embedded in $M$. 

\vspace{2mm}

\noindent
{\it A marked $S$--knot of $M$ an is embedding $\varXi:S\rightarrow M$ of the surface $S$ 
into $M$ such that}
\begin{align}
&\varXi(p_S)=p_M, 
\\
&\varXi\circ\zeta_{Si}=\zeta_{Mi}.
\end{align}

\vspace{2mm}

\noindent
Ambient isotopy is the natural notion of mutual deformability also of 
marked $S$--knots.

\vspace{2mm}

\noindent
{\it Two marked $S$--knots $\varXi_0,\varXi_1$ are ambient isotopic if there is a 
smooth family $F_z\in\Diff_+(M)$, $z\in\mathbbm{R}$, of orientation preserving 
diffeomorphisms such that}
\begin{align}
&F_0=\id_M,\quad \varXi_1=F_1\circ \varXi_0, 
\nonumber
\\
&F_z(p_M)=p_M,~~F_z\circ\zeta_{Mi}=\zeta_{Mi}.
\nonumber
\end{align}

\vspace{2mm}

\noindent
Analogously to $C$--knots, to compute $S$--knot holonomy 
we need parametrized $S$--knots. This is achieved by 
assigning a surface to any marked $S$--knot. 

\vspace{2mm}

\noindent
The fundamental polygon of $S$ is the boundary of the simply connected open $2$--fold that 
results cutting $S$ along the standard $a$-- and $b$--cycles. It plays a basic role
in the subsequent constructions. 

\vspace{2mm}

\noindent
{\it View $S$ as a $C$--marked manifold and let} 
\begin{align}
\gamma_{Si}=\gamma_{\zeta_{Si}} ~~\text{that is}~~
\alpha_{Sr}=\gamma_{\xi_{Sr}}, ~ \beta_{Sr}=\gamma_{\eta_{Sr}},
\end{align}
{\it Then the fundamental polygon of $S$ is the curve given by} 
\begin{align}
&\tau_S=\beta_{S\ell_S}{}^{-1_\circ}\circ\alpha_{S\ell_S}{}^{-1_\circ}\circ\beta_{S\ell_S}\circ
\alpha_{S\ell_S}\circ 
\\
&\hspace{3cm}
\cdots\circ \beta_{S1}{}^{-1_\circ}\circ\alpha_{S1}{}^{-1_\circ}\circ\beta_{S1}\circ\alpha_{S1}
\nonumber
\end{align}
{\it if $\ell_S=0$, $\tau_S=\iota_{p_S}$}. 

\vspace{2mm}

\noindent
As a compatible curve in $C$ is required in order to 
associate a curve to each marked $C$--knot, a compatible surface 
in $S$ is required in order to 
associate a surface to each marked $C$--knot. 

\vspace{2mm}


\noindent
{\it A compatible surface in $S$ is a surface $\varSigma_S:\iota_{p_S}\rightarrow \tau_S$ 
such that
\begin{enumerate}[i)]

\vspace{2mm}
\item
$D_S=\varSigma_S{}^{-1}(S\setminus \cup_i\zeta_{Si}(C))$ is an open simply connected domain in $\mathbbm{R}^2$;

\vspace{2mm}
\item
$\varSigma_S|_{D_S}:D_S\rightarrow S\setminus \cup_i\zeta_{Si}(C)$ is an 
orientation preserving diffeomorphism. 

\end{enumerate}
}

\vspace{2mm}

\noindent
{\it Example}. Let $S=S^2$ be the sphere embedded in $\mathbbm{R}^3$ as 
\begin{align}
&S_{S^2}(\vartheta,\varphi)=(\cos\vartheta\sin\vartheta(1-\cos\varphi),
\end{align}
\begin{align}
&\hspace{3cm}-\sin\vartheta\sin\varphi,1-\sin^2\vartheta(1-\cos\varphi)),
\nonumber
\end{align}
where $\vartheta\in(0,\pi)$, $\varphi\in[0,2\pi)$, with the $S$--marking $p_{S^2}=(0,0,1)$. 
A compatible surface $\varSigma_{S^2}:\mathbbm{R}^2\rightarrow S^2$ is given by
\begin{align}
\varSigma_{S^2}(x,y)=S_{S^2}(\pi\alpha(y),2\pi\alpha(x)),
\end{align}
where $\alpha:\mathbbm{R}\rightarrow [0,1]$ is a function 
enjoying the properties listed below Equation \eqref{gammas1}. 

\vspace{2mm}

\noindent
The surface $\varSigma_{S^2}$ describes a parametrized family of circles on $S^2$ which 
spring from the north pole on one side of it, sweep $S^2$ dilating,  
reaching the south pole and then contracting and finally converge to the north pole 
on the other side. 

\vspace{2mm}

\noindent
{\it Example}. 
Let $S=T^2$ be the torus embedded in $\mathbbm{R}^3$ as
\begin{align}
&S_{T^2}(\vartheta_1,\vartheta_2)=(\cos\vartheta_1(1+r\cos\vartheta_2),
\\
&\hspace{3.5cm}\sin\vartheta_1(1+r\cos\vartheta_2),r\sin\vartheta_2),
\nonumber
\end{align}
where $r<1$ is fixed and $\vartheta_1,\vartheta_2\in[0,2\pi)$, with the 
$S$--marking $p_{T^2}=(1+r,0,0)$ and 
\begin{align}
\xi_{T^2}(\vartheta)&=((1+r)\cos\vartheta,(1+r)\sin\vartheta,0),
\vphantom{\Big]}
\nonumber 
\\
\eta_{T^2}(\vartheta)&=(1+r\cos\vartheta,0,r\sin\vartheta),
\vphantom{\Big]}
\nonumber 
\end{align}
where $\vartheta\in[0,2\pi)$. A compatible surface 
$\varSigma_{T^2}:\mathbbm{R}^2\rightarrow T^2$ is 
\begin{align}
&\varSigma_{T^2}(x,y)=S_{T^2}(2\pi c_1(x,y)),2\pi c_2(x,y)))
\vphantom{\Big]}
\nonumber 
\\
&c_1(x,y)=\varrho(4\alpha(x),\alpha(y))-\varrho(4\alpha(x)-2,\alpha(y)),
\vphantom{\Big]}
\nonumber 
\\
&c_2(x,y)=\varrho(4\alpha(x)-1,\alpha(y))-\varrho(4\alpha(x)-3,\alpha(y)),
\vphantom{\Big]}
\nonumber 
\end{align}
where $\alpha:\mathbbm{R}\rightarrow [0,1]$ is a function 
with the same properties as before and 
$\varrho:\mathbbm{R}\times[0,1]\rightarrow [0,1]$ is the function given by 
\begin{equation}
\varrho(s,t)=tg_\beta\bigg(\frac{1-2s}{(1+s-t)(2-s-t)}\bigg),
\label{sample10}
\end{equation}
where $g_\beta(w)=1/(\exp(\beta w)+1)$ with $\beta>0$ is the Fermi--Dirac function.

\vspace{2mm}

\noindent
Upon unfolding the torus $T^2$ into a square $I^2$ by cutting it along the $a$-- and $b$--cycle,
the surface $\varSigma_{T^2}$ 
describes a parametrized family of closed curves on $I^2$ which spring from one corner
of the square and sweep it all eventually approximating the square's boundary.


\vspace{2mm}

\noindent
A surface furnishing a natural parametrization of a given
marked $S$--knot can now be constructed. To this end, we need 
to identify a curve in $M$ that matches the fundamental 
polygon of $S$. 

\vspace{2mm}

\noindent
{\it View $M$ as a $C$--marked manifold and let}
\begin{align}
\gamma_{Mi}=\gamma_{\zeta_{Mi}} ~~\text{that is}~~
\alpha_{Mr}=\gamma_{\xi_{Mr}}, ~ \beta_{Mr}=\gamma_{\eta_{Mr}},
\end{align}
{\it Then, the fundamental polygon of the marking of $M$ is the curve}
\begin{align}
&\tau_M=\beta_{M\ell_S}{}^{\!-1_\circ}\circ\alpha_{M\ell_S}{}^{\!-1_\circ}\circ\beta_{M\ell_S}\circ\alpha_{M\ell_S}
\circ%
\\
&\hspace{2cm}\cdots\circ
\beta_{M1}{}^{\!-1_\circ}\circ\alpha_{M1}{}^{\!-1_\circ}\circ\beta_{M1}\circ\alpha_{M1}
\nonumber
\end{align}
{\it if $\ell_S=0$, $\tau_M=\iota_{p_M}$.} 
\vspace{2mm}

\noindent
A surface furnishing a natural parametrization of a given marked $S$--knot 
can now be constructed. 

\vspace{2mm}

\noindent
{\it With every marked $S$--knot $\varXi$, there is associated 
a surface $\varSigma_\varXi:\iota_{p_M}\Rightarrow \tau_M$ given by}
\begin{align}
\varSigma_\varXi=\varXi\circ \varSigma_S.
\end{align}

\vspace{2mm}

\noindent
Note that $\tau_M=\varXi\circ\tau_S$. 

\vspace{2mm}

\noindent
For a marked $C$--knot $\xi$, the source and target of the 
associated curve $\gamma_\xi:p_M\rightarrow p_M$ are equal. 
In gauge theory, this ensures nice ambient isotopy and gauge covariance properties 
of $C$--knot holonomy. 
For a genus $\ell_S=0$ marked $S$--knot $\varXi$, the source and target of the 
associated surface $\varSigma_\varXi:\iota_{p_M}\Rightarrow \iota_{p_M}$ are equal as well.
In higher gauge theory, this also ensures nice ambient isotopy and gauge covariance properties 
of $S$--knot holonomy.  
However, for a genus $\ell_S>0$ marked $S$--knot $\varXi$, the source and target of the 
associated surface $\varSigma_\varXi:\iota_{p_M}\Rightarrow \tau_M\not=\iota_{p_M}$ are 
different. This is likely to be a problem for ambient isotopy and gauge covariance 
properties of holonomy. We have a proposal for the solution of this difficulty.

\vspace{2mm}

\noindent
For given $\ell_S$ and $C$--marking of $M$, pick a reference
mar\-ked $S$--knot $\varDelta_M$ (e. g. Hosokawa's and Kawauchi's surface unknots in $S^4$
\cite{hosokawa1979proposals}).

\vspace{2mm}

\noindent
{\it The normalized surface of a marked $S$--knot $\varXi$ is the surface 
$\varSigma^\sharp{}_{\varXi}:\iota_{p_M}\Rightarrow\iota_{p_M}$ given by}
\begin{align}
\varSigma^\sharp{}_{\varXi}=\varSigma_M{}^{-1_\bullet}\bullet\varSigma_{\varXi},
\end{align}
{\it where $\varSigma_M:=\varSigma_{\varDelta_M}$ and $\bullet$ denotes vertical surface 
composition (cf. Section \ref{sec:holo}).}

\vspace{2mm}

\noindent
An intuitive way of thinking of the normalized surface of $\varXi$ is 
as a surface characterizing the $S$--knot ``ratio'' of $\varXi$ to $\varDelta_M$,
with $\varDelta_M$ acting as a normalizing knot. 

\vspace{2mm}

\noindent
The normalized surface of a marked $S$--knot has nice properties. 

\vspace{2mm}

\noindent
{\it $\varSigma^\sharp{}_{\varXi}$ is independent from the choice of $\varSigma_S$ and 
$\gamma_C$ up to thin homotopy.}

\vspace{2mm}

\noindent
Note that the markings $p_C$ and $(p_S,\zeta_{Si})$ are fixed.

\vspace{2mm}

\noindent
Ambient isotopic reference $S$--knots yield homotopic normalized marked $S$--knot surfaces.

\vspace{2mm}

\noindent
{\it If the reference marked $S$--knots 
$\varDelta_{M0}$, $\varDelta_{M1}$ are ambient isotopic, then for every marked $S$--knot $\varXi$
the normalized surfaces $\varSigma^\sharp{}_{\varXi|0}$, $\varSigma^\sharp{}_{\varXi|1}$ are homotopic.}

\vspace{2mm}

\noindent
Ambient isotopic marked $S$--knots have homotopic normalized surfaces. 

\vspace{2mm}

\noindent
{\it If $\varXi_0$, $\varXi_1$ are ambient isotopic marked $S$--knots, then the normalized surfaces 
$\varSigma^\sharp{}_{\varXi_0}$, $\varSigma^\sharp{}_{\varXi_1}$ are homotopic. }

\vspace{2mm}

\noindent
As for $C$--knots, it should be possible to alter the marking changing $\varSigma^\sharp{}_\varXi$ by
a (thin) homotopy.

\vspace{2mm}

\noindent
{\it Two marked $S$--knots $\varXi_0$, $\varXi_1$ with respect to distinct
$S$--markings $(p_{M0},\zeta_{M0i})$, $(p_{M1},\zeta_{M1i})$ of $M$ 
are said to be freely ambient isotopic if there is a 
smooth family $F_z\in\Diff_+(M)$, $z\in\mathbbm{R}$, of orientation preserving 
diffeomorphisms such that}
\begin{align}
F_0=\id_M,\quad \varXi_1=F_1\circ \varXi_0.
\end{align}

\vspace{2mm}

\noindent
Notice that above the $S$--marking $(p_S,\zeta_{Si})$ of $S$ is kept fixed. 

\vspace{2mm}

\noindent
{\it Two pairs $\varXi_0$, $\varXi_1$ and $\varXi_0{}'$, $\varXi_1{}'$ of freely ambient isotopic 
marked $S$--knots are called concordant if there exist ambient isotopies 
$F_z$ of $\varXi_0$, $\varXi_1$ and $F'{}_z$ of 
$\varXi_0{}'$, $\varXi_1{}'$ s. t. $F_z(p_{M0})=F'{}_z(p_{M0})$, 
$F_z\circ\zeta_{M0i}=F'{}_z\circ\zeta_{M0i}$.}

\vspace{2mm}

\noindent
Freely ambient isotopic marked $S$--knots have homotopic normalized surfaces up to conjugation
under concordance with reference knots.

\vspace{2mm}

\noindent
{\it  Suppose the reference marked $S$--knots $\varDelta_{M0}$, $\varDelta_{M1}$ are 
freely ambient isotopic. If the marked $S$--knots $\varXi_0$, $\varXi_1$ are freely ambient 
isotopic concordantly with $\varDelta_{M0}$, $\varDelta_{M1}$, then there is curve a
$\gamma_1:p_{M0}\rightarrow p_{M1}$ such that
$\varSigma^\sharp{}_{\varXi_0|0}$, $I_{\gamma_1}{}^{-1_\circ}
\circ\varSigma^\sharp{}_{\varXi_1|1}\circ I_{\gamma_1}$ are homotopic.}

\vspace{2mm}

\noindent
Before stating the next result, we recall the following 
property. For two $S$--markings $(p_{S0},\zeta_{S0i})$, $(p_{S1},\zeta_{S1i})$
of $S$, there is an orientation preserving ambient isotopy $k_z$ of $S$ such that
$k_1(p_{S0})=p_{S1}$, $k_1\circ \zeta_{S0i}=\zeta_{S1i}$.

\vspace{2mm}

\noindent
{\it If the embeddings $\varDelta_M,\varXi:S\rightarrow M$ are simultaneously the 
reference and considered marked $S$--knot with respect to two distinct $S$--markings 
$(p_{S0},\zeta_{S0i})$, $(p_{M0},\zeta_{M0i})$ and $(p_{S1},\zeta_{S1i})$, $(p_{M1},\zeta_{M1i})$ 
of $S$ and $M$ and there is an ambient isotopy $k_z$ of $S$ 
shifting $\{p_{S0},\zeta_{S0i}\}$ to $\{p_{S1},\zeta_{S1i}\}$ 
such that $\varXi\circ k_z(p_{S0})=\varDelta_M\circ k_z(p_{S0})$ and 
$\varXi\circ k_z\circ \zeta_{S0i}=\varDelta_M\circ k_z\circ \zeta_{S0i}$, then there is 
a curve $\gamma_1:p_{M0}\rightarrow p_{M1}$ lying in the image $\varXi(S)$ such that
$\varSigma^\sharp{}_{\varXi|0}$, $I_{\gamma_1}{}^{-1_\circ}\circ\varSigma^\sharp{}_{\varXi|1}\circ I_{\gamma_1}$
are thinly homotopic.}

\vspace{2mm}

Relying on the above results, we can now tackle the task of 
constructing higher knot holonomy.

\section{$C$-- and $S$--knot holonomy}\label{sec:csknothol}

Our aim is constructing holonomy invariants of knots up to conjugation.
We begin with reviewing how this is done for $C$--knots

\vspace{2mm}

\noindent
We let $G$ be a Lie group and $\theta$ be a flat $G$--connection on $M$.
Further, we fix $C$--markings $p_C$ and $p_M$ of $C$ and $M$, respectively.

\vspace{2mm}

\noindent
The holonomy of a marked $C$--knot is built out of the curve associated
to the knot.

\vspace{2mm}

\noindent
{\it The holonomy of a marked $C$--knot $\xi$ is the element $F_\theta(\xi)\in G$ given by }
\begin{align}
F_\theta(\xi)=F_\theta(\gamma_\xi),
\end{align}
{\it where $\gamma_\xi:p_M\rightarrow p_M$ curve of $\xi$ (cf. Section \ref{sec:csknots})
and $F_\theta$ is the parallel transport functor (cf. Section \ref{sec:partr}).}

\vspace{2mm}

\noindent
$C$--knot holonomy is independent from the choice of parametrization.

\vspace{2mm}

\noindent
{\it For any marked $C$--knot $\xi$, $F_\theta(\xi)$
is independent of the choice of the compatible curve $\gamma_C$ of $C$.}

\vspace{2mm}

\noindent
$C$--knot holonomy is further invariant under ambient isotopy. 

\vspace{2mm}

\noindent
{\it If $\xi_0$, $\xi_1$ are ambient isotopic marked $C$--knots of $M$, then }
\begin{align}
F_\theta(\xi_1)=F_\theta(\xi_0). 
\end{align}

\vspace{2mm}

\noindent
This property generalizes as follows. Fix the $C$--marking $p_C$ of $C$ but allow two 
distinct $C$--markings $p_{M0}$, $p_{M1}$ of $M$.

\vspace{2mm}

\noindent
{\it If $\xi_0$, $\xi_1$ are freely ambient isotopic marked $C$--knots, then 
there exists a curve $\gamma_1:p_{M0}\rightarrow p_{M1}$ of $M$ such that  }
\begin{align}
F_\theta(\xi_1)=F_\theta(\gamma_1)F_\theta(\xi_0)F_\theta(\gamma_1)^{-1}.
\nonumber 
\end{align}

\vspace{2mm}

\noindent
$C$--knot holonomy is independent of the way a given $C$--knot is marked
up to conjugation. 

\vspace{2mm}

\noindent
{\it If $\xi$ is a marked $C$--knot with respect to two 
distinct $C$--markings $p_{C0}$, $p_{M0}$ and $p_{C1}$, $p_{M1}$ of $C$ and $M$, then
there is a curve $\gamma_1:p_{M0}\rightarrow p_{M1}$ lying in $\xi(C)$ such that  }
\begin{align}
F_{\theta|1}(\xi)=F_\theta(\gamma_1)F_{\theta|0}(\xi)F_\theta(\gamma_1)^{-1}.
\nonumber 
\end{align}

\vspace{2mm}

\noindent
$C$--knot holonomy is also gauge covariant as desired. 

\vspace{2mm}

\noindent
{\it Let $\xi$ be a marked $C$--knot of $M$. Then, for any $G$--gauge transformation $g$, one has}
\begin{align}
F_{{}^g\theta}(\xi)=g(p_M)F_\theta(\xi)g(p_M)^{-1}.
\nonumber 
\end{align}

\vspace{2mm}

\noindent
In summary, $C$--knot holonomy is $C$--marking and gauge independent and isotopy invariant 
up to $G$--conjugation.

\vspace{2mm}


Next, using the treatment of $C$--knot holonomy presented above as a model,
we illustrate the construction of $S$--knot holonomy.

\vspace{2mm}

\noindent
We let $(G,H)$ be a Lie crossed module and $(\theta,\varUpsilon)$ be 
a flat $(G,H)$--$2$--connection pair on $M$. Furthermore, 
we fix $S$-markings $(p_S,\zeta_{Si})$ and $(p_M,\zeta_{SMi})$ of $S$ and $M$, respectively.

\vspace{2mm}

\noindent
{\it The holonomy of a marked $S$--knot $\varXi$ is the element $F_\theta(\varXi)$ $\in H$ given by}
\begin{align}
F_{\theta,\varUpsilon}(\varXi)=F_{\theta,\varUpsilon}(\varSigma^\sharp{}_\varXi)
=F_{\theta,\varUpsilon}(\varSigma_M)^{-1}F_{\theta,\varUpsilon}(\varSigma_\varXi),
\end{align}
{\it where $\varSigma^\sharp{}_\varXi:\iota_{p_M}\Rightarrow \iota_{p_M}$ is the 
normalized surface of $\varXi$ and $F_{\theta,\varUpsilon}$ is the parallel transport 2--functor.}

\vspace{2mm}

\noindent
The fact that $\varSigma^\sharp{}_\varXi:\iota_{p_M}\Rightarrow \iota_{p_M}$
has the following crucial consequence. 

\vspace{2mm}

\noindent
{\it For a marked $S$--knot $\varXi$, } 
\begin{align}
t(F_{\theta,\varUpsilon}(\varXi))=1_G.
\end{align} 
{\it Thus, $F_{\theta,\varUpsilon}(\varXi)=1_H$ unless $\ker t\not= \{1_H\}$. 
Further, $F_{\theta,\varUpsilon}(\varXi)\in Z_H$.}

\vspace{2mm}

\noindent
Thus, unlike $C$--knot holonomy, $S$--knot holonomy is fundamentally Abelian 
and non trivial only for crossed modules whose target map has non trivial kernel. 

\vspace{2mm}

\noindent
$S$--knot holonomy is independent from the choice of parametrization.

\vspace{2mm}

\noindent
{\it For every marked $S$--knot $\varXi$, $F_{\theta,\varUpsilon}(\varXi)$ is 
independent from the choice of the compatible surface $\varSigma_S$ of $S$ and curve 
$\gamma_C$ of $C$.}

\vspace{2mm}

\noindent
$S$--knot holonomy is invariant under a change of the reference marked $S$--knots 
in the following sense. 

\vspace{2mm}

\noindent
{\it If the reference marked $S$--knots $\varDelta_{M0}$, $\varDelta_{M1}$ are ambient isotopic, then
for any marked $S$--knot $\varXi$}
\begin{align}
F_{\theta,\varUpsilon|0}(\varXi)=F_{\theta,\varUpsilon|1}(\varXi).
\end{align}

\vspace{2mm}

\noindent
$S$--knot holonomy is further invariant under ambient isotopy. 

\vspace{2mm}

\noindent
{\it If $\varXi_0$, $\varXi_1$ are ambient isotopic marked $S$--knots, then }
\begin{align}
F_{{\theta,\varUpsilon}}(\varXi_1)=F_{\theta,\varUpsilon}(\varXi_0).
\end{align}

\vspace{2mm}

\noindent
This property generalizes as follows. Fix the $S$-markings $(p_S,\zeta_{Si})$ of $S$ but allow 
distinct $S$--marking $(p_{M0},\zeta_{M0i})$, $(p_{M1},\zeta_{M1i})$ of $M$.

\vspace{2mm}

\noindent
{\it Suppose $\varDelta_{M0}$, $\varDelta_{M1}$ are freely ambient isotopic reference marked $S$--knots. 
If the marked $S$--knots $\varXi_0$, $\varXi_1$ are freely ambient isotopic concordantly with 
$\varDelta_{M0}$, $\varDelta_{M1}$, then there is a curve $\gamma_1:p_{M0}\rightarrow p_{M1}$ such that}
\begin{align}
F_{\theta,\varUpsilon|1}(\varXi_1)=m(F_\theta(\gamma_1))(F_{\theta,\varUpsilon|0}(\varXi_0)).
\end{align}

\vspace{2mm}

\noindent
$S$--knot holonomy is independent of the way a given knot $S$--knot is marked
up to conjugation. 

\vspace{2mm}

\noindent
{\it If the embeddings $\varDelta_M,\varXi:S\rightarrow M$ are simultaneously the 
reference and considered marked $S$--knot with respect to two distinct $S$--markings 
$(p_{S0},\zeta_{S0i})$, $(p_{M0},\zeta_{M0i})$ and $(p_{S1},\zeta_{S1i})$, $(p_{M1},\zeta_{M1i})$ 
of $S$ and $M$ and there is an ambient isotopy $k_z$ of $S$ 
shifting $\{p_{S0},\zeta_{S0i}\}$ to $\{p_{S1},\zeta_{S1i}\}$ 
such that $\varXi\circ k_z(p_{S0})=\varDelta_M\circ k_z(p_{S0})$ and 
$\varXi\circ k_z\circ \zeta_{S0i}=\varDelta_M\circ k_z\circ \zeta_{S0i}$, then there is 
a curve $\gamma_1:p_{M0}\rightarrow p_{M1}$ lying in the image $\varXi(S)$ such that}
\begin{align}
F_{\theta,\varUpsilon|1}(\varXi)=m(F_\theta(\gamma_1))(F_{\theta,\varUpsilon|0}(\varXi)).
\end{align}

\vspace{2mm}

\noindent
In this higher gauge theoretic set--up, one can define also $C$--knot holonomy
in the same way as before. 

\vspace{2mm}

\noindent
{\it The holonomy of a marked $C$--knot $\xi$ is the element $F_\theta(\xi)\in G$ given by }
\begin{align}
F_\theta(\xi)=F_\theta(\gamma_\xi)
\end{align}
{\it is defined.}

\vspace{2mm}

\noindent
This $C$--knot holonomy has however weaker properties than in ordinary gauge theory. 

\vspace{2mm}

\noindent
$C$--knot holonomy is still independent of the choice of parametrization.

\vspace{2mm}

\noindent
{\it For any marked $C$--knot $\xi$, $F_\theta(\xi)$
is independent of the choice of the compatible curve $\gamma_C$ of $C$.}

\vspace{2mm}

\noindent
Since, however, $\theta$ is not flat unless $\dot t(\varUpsilon)=0$, $F_\theta(\xi)$ 
is not ambient isotopy invariant.

\vspace{2mm}

\noindent
{\it If $\xi_0$, $\xi_1$ are two ambient isotopic marked $C$--knots of $M$, then there is 
a surface $\varSigma:\gamma_{\xi_0}\Rightarrow \gamma_{\xi_1}$ of $M$ such that }
\begin{align}
F_\theta(\xi_1)=t(F_{\theta,\varUpsilon}(\varSigma))F_\theta(\xi_0). 
\end{align}

\vspace{2mm}

\noindent
This property generalizes as follows. Fix the $C$--marking $p_C$ of $C$ but allow two 
distinct $C$--markings $p_{M0}$, $p_{M1}$ of $M$.

\vspace{2mm}

\noindent
{\it If $\xi_0$, $\xi_1$ are freely ambient isotopic marked $C$--knots, then 
there exist a curve $\gamma_1:p_{M0}\rightarrow p_{M1}$ and a surface  
$\varSigma:\gamma_{\xi_0}\Rightarrow\gamma_1{}^{-1_\circ }\circ\gamma_{\xi_1}\circ\gamma_1$ of $M$ such that}
\begin{align}
F_\theta(\xi_1)=F_\theta(\gamma_1)t(F_{\theta,\varUpsilon}(\varSigma))
F_\theta(\xi_0)F_\theta(\gamma_1)^{-1}.
\end{align}

\vspace{2mm}

\noindent
$C$--knot holonomy is again independent of the way a given $C$--knot is marked
up to conjugation. 

\vspace{2mm}

\noindent
{\it If $\xi$ is a marked $C$--knot with respect to two 
distinct $C$--markings $p_{C0}$, $p_{M0}$ and $p_{C1}$, $p_{M1}$ of $C$ and $M$, then
there is $\gamma_1:p_{M0}\rightarrow p_{M1}$ curve in $\xi(C)$ such that  }
\begin{align}
F_{\theta|1}(\xi)=F_\theta(\gamma_1)F_{\theta|0}(\xi)F_\theta(\gamma_1)^{-1}.
\label{xiholo2}
\end{align}

\vspace{2mm}

\noindent
Also in higher gauge theory, $S$-- and $C$--knot holonomy is gauge covariant in the appropriate sense. 

\vspace{2mm}

\noindent
{\it Let $\varXi$ be a marked $S$--knot and $\xi$ a marked $C$--knot. 
If $(g,J)$ is a $(G,H)$--$1$--gauge transformation, then}
\begin{align}
F_{{}^{g,J}\theta,{}^{g,J}\varUpsilon}(\varXi)=m(g(p_M))(F_{\theta,\varUpsilon}(\varXi))
\end{align}
{\it and } 
\begin{align}
F_{{}^{g,J}\theta}(\xi)=g(p_M)t(G_{g,J;\theta}(\gamma_\xi))F_\theta(\gamma)g(p_M)^{-1},
\end{align}
{\it where $G_{g,J;\theta}(\gamma_\xi)$ is the gauge parallel transport along $\gamma_\xi$
defined in Section \ref{sec:partr}.}

\vspace{2mm}

To summarize, $C$-- and $S$--knot holonomy are $C$--mar\-king and gauge independent
and isotopy invariant up to the appropriate form of crossed  module conjugation.
We shall analyze this point in greater depth in the next section.

\section{Invariant traces} \label{sec:invtr}

Having applications to knot topology in mind, we aim at a construction 
of holonomy invariants. This requires working out invariant traces.

\vspace{2mm}

\noindent
We let $G$ again be a Lie group and $\theta$ be a flat $G$--connec\-tion on $M$.
Further, we let $C$--markings $p_C$ and $p_M$ of $C$ and $M$, respectively, be given.

\vspace{2mm}

\noindent
We have seen in Section \ref{sec:csknothol} that for a $C$--knot $\xi$, 
its holonomy $F_\theta(\xi)$ is $C$--marking and isotopy invariant and gauge 
independent up to $G$--conjugation, that is 
\begin{align}
F_\theta(\xi) \equiv a F_\theta(\xi) a^{-1}
\end{align}
for $a\in G$.

\vspace{2mm}

\noindent
There is a well established way of extracting $C$--knot invariants from 
knot holonomy. 

\vspace{2mm}

\noindent
{ \it The Wilson line}
\begin{align}
W_{R,\theta}(\xi)=\tr_R(F_\theta(\xi)),
\end{align}
{\it with $R$ a representation of $G$ provides a $C$--knot invariant.}


\vspace{2mm}

Next, taking the procedure just reviewed to construct $C$--knot holonomy 
invariants as a model, we propose a systematic way to build $S$--knot holonomy
invariants.

\vspace{2mm}

\noindent
We let $(G,H)$ be a Lie crossed module and $(\theta,\varUpsilon)$ be 
a flat $(G,H)$--$2$--connection pair on $M$. Furthermore, 
we let $S$-markings $(p_S,\zeta_{Si})$ and $(p_M,\zeta_{SMi})$ of $S$ and $M$, 
respectively, be given.

\vspace{2mm}

\noindent
In Section \ref{sec:csknothol}, we have also seen that for a $C$--knot $\xi$ and an 
$S$--knot $\varXi$, the holonomy $F_\theta(\xi)$ 
and $F_{\theta,\Upsilon}(\varXi)$ is $C$-- and 
$S$--marking and isotopy invariant and gauge independent up to $(G,H)$--conjugation
\begin{align}
&F_\theta(\xi) \equiv a F_\theta(\xi) a^{-1}t(A),
\\
&F_{\theta,\Upsilon}(\varXi)\equiv m(a)(F_{\theta,\Upsilon}(\varXi))
\end{align}
with $(a,A)\in G\times H$. $(G,H)$--conjugation is defined by 
\begin{align}
u'=aua^{-1}t(A),\qquad U'=m(a)(U)
\end{align}
with $(u,U),(u',U'),(a,A)\in G\times H$ and is an equivalence relation.

\vspace{2mm}

\noindent
To obtain knot invariants, one needs traces invariant under $(G,H)$--conjugation.
To this end, one could proceed as follows.

\vspace{2mm}

\noindent
Assume $G$, $H$ are compact with bi-invariant Haar measures $\mu_G$, $\mu_H$.
Pick representations $R$, $S$ of $G$, $H$. Set  
\begin{align}
&\tr_{R,S|b}(u)=\int_Hd\mu_H(X)\tr_R(ut(X)),
\label{trb}
\\
&\tr_{R,S|f}(U)=\int_Gd\mu_G(x)\tr_S(m(x)(U)),
\label{trf}
\end{align}
$(u,U)\in G\times H$. 

\vspace{2mm}

\noindent
{\it The traces $\tr_{R,S|b}$, $\tr_{R,S|f}(U)$ are invariant under $(G,H)$ conjugation},
\begin{align}
&\tr_{R,S|b}(aua^{-1}t(A))=\tr_{R,S|b}(u),
\\
&\tr_{R,S|f}(m(a)(U))=\tr_{R,S|f}(U)
\end{align}
{\it for $(u,U), (a,A)\in G\times H$}. 

\vspace{2mm}

\noindent
These invariant traces can be used to extract $C$-- and $S$--knot invariants from 
knot holonomy as in the ordinary case. 

\vspace{2mm}

\noindent
{\it The Wilson line and surface}
\begin{align}
&W_{R,S,\theta|b}(\xi)=\tr_{R,S|b}(F_\theta(\xi)),
\\
&W_{R,S,\theta,\Upsilon|f}(\varXi)=\tr_{R,S|f}(F_{\theta,\Upsilon}(\varXi)). 
\end{align}
{\it provide a $C$-- and $S$--knot invariant.}

\vspace{2mm}

\noindent
There is a problem with this way of proceeding.
The traces may be trivial. For instance, if $t(H)=G$, $\tr_{R,Sb}(u)$ does not depend 
on $u$ and $\tr_{R,Sf}(U)=\tr_S(U)$ for $U\in \ker t$ (the case of interest for surface knots). 


\vspace{2mm}

\noindent
In ordinary gauge theory with gauge group $G$, a trace is a map $\tr:G\rightarrow \mathbbm{C}$ 
invariant under the action 
\begin{align}
a\varrhd u:=aua^{-1}
\end{align}
with $a,u\in G$, that is 
\begin{align}
\tr(a\varrhd u)=\tr(u). 
\end{align}
If $G$ is a compact Lie group, then $\tr$ reduces to a linear combination
of ordinary traces $\tr_R$ associated with the irreducible representations $R$ 
of $G$.
\vspace{2mm}

\noindent
What matters is not the group structure of $G$ but its conjugation structure 
codified in the conjugation pointed quandle of $G$. 

\vspace{2mm}

\noindent
{\it A pointed quandle is a set $G$ with a binary operation $\varrhd:G\times G\rightarrow G$ and 
a distinguished element $1_G\in G$ such that}
\begin{align}
&a\varrhd a=a,
\\
&a\varrhd(b\varrhd c)=(a\varrhd b)\varrhd(a\varrhd c)
\end{align}
{\it with $a,b,c\in G$. Further, the map 
$a\varrhd\cdot:G\rightarrow G$ is invertible for any $a\in G$ and}
\begin{align}
a\varrhd 1_G=1_G,\quad 1_G\varrhd a=a
\end{align}
{\it for $a\in G$}.

\vspace{2mm}

In higher gauge theory with gauge crossed module $(G,H)$, 
a similar point of view is appropriate. A trace pair is a pair of maps 
$\tr_b:G\rightarrow \mathbbm{C}$, $\tr_f:H\rightarrow \mathbbm{C}$ invariant under
the action 
\begin{align}
&a\varrhd u:=aua^{-1}~,
\\
&A\varsucc u:=ut(A)~,
\\
&a\varrhd U:=m(a)(U)
\end{align}
with $a,u\in G$, $A,U\in H$, that is 
\begin{align}
&\tr_b(a\varrhd u)=\tr_b(u),
\\
&\tr_b(A\varsucc u)=\tr_b(u),
\\
&\tr_f(a\varrhd U)=\tr_f(U).
\end{align}

\vspace{2mm}

\noindent
What matters is not $(G,H)$ itself but its conjugation augmented pointed quandle crossed module $(G,H)$
\cite{Crans:2004ve,Crans:1310.4705,Zucchini:2015xba}: 

\vspace{2mm}

\noindent
{\it An augmented pointed quandle crossed module is a pair of sets $G$, $H$ endowed with three operations 
$\varrhd:G\times G\rightarrow G$, $H\times H\rightarrow H$, $G\times H\rightarrow H$ and distinguished 
elements $1_G\in G$, $1_H\in H$ such that 
\begin{enumerate}[i)]

\vspace{2mm}
\item
$G$ is a pointed quandle,

\vspace{2mm}
\item
$H$ is a pointed quandle

\end{enumerate}

\vspace{2mm}

\noindent
and the following requirements are satisfied.

\vspace{2mm}

\noindent
The relations
\begin{align}
&a\varrhd(b\varrhd A)=(a\varrhd b)\varrhd(a\varrhd A), 
\\
&a\varrhd(A\varrhd B)=(a\varrhd A)\varrhd(a\varrhd B)
\end{align}
with $a,b\in G$, $A,B\in H$ hold. 

\vspace{2mm}

\noindent  
For any $a\in G$, the map $a\varrhd \cdot:H\rightarrow H$ is invertible.

\vspace{2mm}

\noindent  
For $a\in G$, $A\in H$, the relations
\begin{align}
&1_G\varrhd A=A,
\\
&a\varrhd 1_H=1_H
\end{align}
are fulfilled.

\vspace{2mm}

\noindent
Further, a quandle morphism $\alpha:H\rightarrow G$ (a map respecting $\varrhd$ and $1$) is given
such that  
\begin{align}
&\alpha(a\varrhd A)=a\varrhd \alpha(A)~,
\label{trg}
\\
&\alpha(A)\varrhd B=A\varrhd B
\label{peif}
\end{align}
with $a\in G$, $A,B\in H$.

\vspace{2mm}

\noindent
Finally, an augmentation map $\varsucc:H\times G\rightarrow G$ is given with the following 
properties.

\vspace{2mm}

\noindent
For $a,b\in G$, $A\in H$, 
\begin{align}
a\varrhd(A\varsucc b)=(a\varrhd A)\varsucc(a\varrhd b)~.
\end{align}

\vspace{2mm}

\noindent
For $A\in H$, $A\varsucc \cdot:G\rightarrow G$ is invertible.

\vspace{2mm}

\noindent
For $a\in G,~A\in H$
\begin{align}
&A\varsucc 1_G=\alpha(A),
\\
&1_H\varsucc a=a.
\end{align}
}

\vspace{2mm}

\noindent
Notice that $\alpha$ is the quandle crossed module analog of the group 
crossed module target morphisms. In particular \eqref{peif} is the quandle counterpart
of the Peiffer identity.

\vspace{2mm}

\noindent
The following question is still open.  
If $G$, $H$ are compact, does a trace pair $\tr_b$, $\tr_f$ reduce to 
linear combinations of traces $\tr_{R,S|b}$, $\tr_{R,S|f}$ of the form \eqref{trb}, 
\eqref{trf} with $R$, $S$ irreducible representations 
of $G$, $H$, respectively?

\section{Higher Chern--Simons theory}

To compute knot invariants in quantum field theory, one needs Chern--Simons theory.
This has been known for a long time since Witten's 1988 paper \cite{Witten:1988hf}.

\vspace{2mm}

\noindent
Chern--Simons theory is a Schwarz type topological gau\-ge theory
on a closed 3-dimensional manifold $M_3$. Suppose that $G$ is the gauge group
and $\mathfrak{g}$ is its Lie algebra.
Suppose further that $\mathfrak{g}$ is equipped with a 
properly normalized invariant non singular bilinear form 
$(\cdot,\cdot):\mathfrak{g}\times\mathfrak{g}\rightarrow \mathfrak{g}$
so that 
\begin{align}
([z,x],y)+x,[z,y])=0
\end{align}
with $x,y,z\in \mathfrak{g}$. 

\vspace{2mm}

\noindent
{\it The Chern--Simons action is given by}
\begin{align}
\mathrm{CS}(\theta)=\frac{k}{4\pi}\int_{M_3}\bigg(\theta,{\rm d}\theta+\frac{1}{3}[\theta,\theta]\bigg)
\end{align}
{\it with $\theta$ a $G$--connection.} The coefficient $k$ is called level. 

\vspace{2mm}

\noindent
{\it The Chern--Simons field equations are equivalent to the flatness condition of $\theta$
(cf. Equation \eqref{twoholo7}):}
\begin{align}
{\rm d}\theta+\frac{1}{2}[\theta,\theta]=0.
\end{align}

\vspace{2mm}

\noindent
{\it The Chern--Simons action is invariant under a $G$--gauge transformations $g$ only
modulo $2\pi\mathbbm{Z}$:}
\begin{align}
\mathrm{CS}({}^g\theta)=\mathrm{CS}(\theta)-2\pi k\cdot\mathrm{wn}(g), 
\end{align}
{\it where $\mathrm{wn}(g)$ is the winding number of $g$.}

\vspace{2mm}

\noindent
{\it Quantum gauge invariance holds if the level $k$ is integer.}

\vspace{2mm}

\noindent
Chern--Simons correlators of Wilson loop $W_{R,\theta}(\xi)$ yield 
knot invariants, for instance:

\vspace{2mm}
$G=\SU(2)$, $R=F$ $\Rightarrow$ Jones polynomial;

\vspace{2mm}
$G=\SU(n)$, $R=F$ $\Rightarrow$ HOMFLY polynomial;

\vspace{2mm}
$G=\SO(n)$, $R=F$ $\Rightarrow$ Kauffman polynomial...

\vspace{2mm}

\noindent
In the Chern--Simons path integral, $\theta$ is not flat and consequently 
$W_{R,\theta}(\xi)$ is not ambient isotopy invariant. However, the theory somehow 
localizes on the moduli space of flat connections even though it is not a 
cohomological topological field theory. This has been proven by Beasley and Witten 
for $M_3$ Seifert, e. g. $S^1\times S^2$, $S^3$, \ldots. 
Therefore, Chern--Simons Wilson loop correlators $W_{R,\theta}(\xi)$ 
furnish genuine knot invariants.

\vspace{2mm}
In order to compute surface knots invariants in quantum field theory, 
one needs a higher version of Chern--Simons theory, 
2-Chern--Simons theory. We have a proposal for such a model.
There are however unsolved problems to be discussed below. 

\vspace{2mm}

\noindent
2--Chern--Simons theory is a Schwarz type topological gauge theory
on a closed 4-dimensional manifold $M_4$. Assume that 
$H\longrightarrow\!\!\!\!\!\!\!\!\!\!\!{}^t\,\,\,\,\,\,\,G
\longrightarrow\!\!\!\!\!\!\!\!\!\!\!\!\!{}^m\,\,\,\,\Aut(H)$ 
is the gauge Lie crossed module and that 
$\mathfrak{h}\longrightarrow\!\!\!\!\!\!\!\!\!\!\!{}^{\dot t}\,\,\,\,\,\,\,\mathfrak{g}
\longrightarrow\!\!\!\!\!\!\!\!\!\!\!\!\!\!{}^{\widehat m}\,\,\,\,\,\,\mathfrak{der}(\mathfrak{h})$
is its differential Lie crossed module. Assume further $(\mathfrak{g},\mathfrak{h})$ is 
equipped with a properly normalized invariant non singular bilinear pairing $(\cdot,\cdot):
\mathfrak{g}\times\mathfrak{h}\rightarrow \mathbbm{R}$ such that 
\begin{align}
&(\dot t(X),Y)-(\dot t(Y),X)=0, 
\nonumber
\\
&([y,x],X)+(x,\widehat{m}(y)(X))=0
\nonumber
\end{align}
with $x,y\in\mathfrak{g}$, $X,Y\in\mathfrak{h}$. 
Note that this requires that the crossed module is balanced, that is
$\dim\mathfrak{g}=\dim\mathfrak{h}$.

\vspace{2mm}

\noindent
{\it The 2--Chern--Simons action is given by}
\begin{align}
\mathrm{CS}_2(\theta,\Upsilon)
=\kappa_2\int_{M_4} \bigg({\rm d}\theta+\frac{1}{2}[\theta,\theta]-\frac{1}{2}\dot t(\Upsilon),\Upsilon\bigg),
\end{align}
{\it with $(\theta,\Upsilon)$ a $(G,H)$ $2$--preconnection \cite{Soncini:2014ara,Zucchini:2015ohw}. 
$\kappa_2$ is a coefficient analog to level.}

\vspace{2mm}

\noindent
A $(G,H)$ $2$--preconnection is just a pair 
$(\theta,\Upsilon)\in\Omega^1(M_4,\mathfrak{g})\times\Omega^2(M_4,\mathfrak{h})$. 
$(\theta,\Upsilon)$ a $(G,H)$--2--connection if in addition it satisfies  
the vanishing fake curvature condition \eqref{twoholo8}.

\vspace{2mm}

\noindent
{\it The $2$--Chern--Simons field equations are equivalent to $(\theta,\Upsilon)$ 
being a flat $(G,H)$--2--connection (cf. Equation \eqref{twoholo8}), \eqref{twoholo27}):}
\begin{align}
&{\rm d}\theta+\frac{1}{2}[\theta,\theta]-\dot t(\varUpsilon)=0,
\label{}
\\
&{\rm d}\varUpsilon+\widehat{m}(\theta,\varUpsilon)=0.
\label{}
\end{align}
{\it Thus, at once, $(\theta,\Upsilon)$ satisfies the vanishing fake curvature condition,
which makes it a genuine $(G,H)$--2--connection, and the vanishing curvature condition, 
which characterizes as a flat one}.

\vspace{2mm}

\noindent
This is quite nice, but it signals a potential problem for the construction of 
2--Chern--Simons theory as a full quantum field theory. Since $(\theta,\Upsilon)$ does not 
obey the zero fake curvature condition in the 2--Chern--Simons path integral,
the insertion of Wilson surfaces $W_{R,S,\theta,\Upsilon}(\varXi)$ 
of surface knots $\varXi$ in the path integral is problematic, as the definition of the 
$W_{R,S,\theta,\Upsilon}(\varXi)$ requires that condition in a basic way.

\vspace{2mm}

\noindent
Another unexpected feature of the model concerns $1$--gauge invariance. 

\vspace{2mm}

\noindent
{\it The 2-Chern--Simons action is invariant under $(G,H)$--1--gauge transformation $(g,J)$,}
\begin{align}
\mathrm{CS}_2({}^{g,J}\theta,{}^{g,J}\theta\Upsilon)=\mathrm{CS}_2(\theta,\Upsilon).
\label{}
\end{align}

\vspace{2mm}

\noindent
In $2$--Chern--Simons theory, there is no shift by some kind of higher winding number such to cause 
level quantization as in the ordinary Chern--Simons model. This surprising and somewhat disappointing 
finding can be explained by hypothesising that either all $(G,H)$--1--gauge transformations $(g,J)$ are small 
unlike ordinary gauge transformation or that we are missing all the topologically non trivial 
$(G,H)$--1--gauge transformations. This is still an open problem. 

\vspace{2mm}

\noindent
In spite of these open issues, the possibility of obtaining 
surface knot invariants as correlators of Wilson surface insertion in 
2--Chern--Simons theory remains an intriguing possibility. 
Here are further reasons for this. 

\vspace{2mm}

\noindent
Studying pull--backs of knots may be interesting.

\vspace{2mm}

\noindent
All orientation preserving diffeomorphisms $f\in\Diff_+(C)$ of the circle $C$ are homotopic to $\id_C$.
Consequently, for a $C$--knot $\xi$, the curves $\gamma_\xi$, $\gamma_{f^*\xi}$ are thinly homotopic\
and the $C$--knots $\xi$ and $f^*\xi$ have the same holonomy.

\vspace{2mm}

\noindent
Conversely, for a higher genus surface $S$, not all orientation 
preserving diffeomorphisms $f\in\Diff_+(S)$ of $S$ 
are homotopic to $\id_S$. Consequently, for a $S$--knot $\varXi$, the normalized 
surfaces $\varSigma^\sharp{}_\varXi$, $\varSigma^\sharp{}_{f^*\varXi}$ are not thinly homotopic
and the $S$--knots $\varXi$ and $f^*\varXi$ do not have the same holonomy in general

\vspace{2mm}

\noindent
This suggests that $S$--knot invariants computed using higher gauge theory may have 
interesting covariance properties under the mapping class group 
\begin{align}
\text{MCG}_+(S)=\Diff_+(S)/\Diff_0(S),
\end{align}
about which there exists a well--developed mathematical theory.

\begin{acknowledgements}
We thank R. Picken for his interest in the subject and for
many useful discussions. 
\end{acknowledgements}

\bibliography{allbibtex}

\begin{thebibliography}{10}

\bibitem{Kauffman:1991ds}
L.~H.~Kauffman,
{\em {Knots and physics},} World Scientific, Singapore, 1991.

\bibitem{adams2004knot}
C.~C.~Adams,
{\em The knot book: an elementary introduction to the mathematical theory of
  knots,} American Mathematical Society, Providence, 2004.

\bibitem{lickorish1997introduction}
W.~R.~Lickorish,
{\em An introduction to knot theory,} Springer, 1997.

\bibitem{carter1998knotted}
J.~S.~Carter and M.~Saito,
{\em Knotted surfaces and their diagrams,} American Mathematical Society,
  Providence, 1998.

\bibitem{kamada2002braid}
S.~Kamada,
{\em Braid and knot theory in dimension four,} American Mathematical Society,
  Providence, 2002.

\bibitem{Wilson:1974sk}
K.~G.~Wilson,
{\em {Confinement of quarks},}
\href{http://dx.doi.org/10.1103/PhysRevD.10.2445}{Phys. Rev. D {\bf 10}  (1974)
  2445}.

\bibitem{Witten:1988hf}
E.~Witten,
{\em {Quantum field theory and the Jones polynomial},}
\href{http://dx.doi.org/10.1007/BF01217730}{Commun. Math. Phys. {\bf 121}
  (1989) 351}.

\bibitem{Chepelev:2001mg}
I.~Chepelev,
{\em Non-Abelian Wilson surfaces,}
\href{http://dx.doi.org/10.1088/1126-6708/2002/02/013}{JHEP {\bf 0202}  (2002)
  013} [{\tt \href{http://www.arxiv.org/abs/hep-th/0111018}{hep-th/0111018}}].

\bibitem{Alekseev:2015hda}
A.~Alekseev, O.~Chekeres, and P.~Mnev,
{\em {Wilson surface observables from equivariant cohomology},}
\href{http://dx.doi.org/10.1007/JHEP11(2015)093}{JHEP {\bf 1511}  (2015) 093}
  [{\tt \href{http://www.arxiv.org/abs/1507.06343}{1507.06343 [hep-th]}}].

\bibitem{Baez:2002jn}
J.~C.~Baez,
{\em {Higher Yang--Mills theory},}
{\tt \href{http://www.arxiv.org/abs/hep-th/0206130}{hep-th/0206130}}.

\bibitem{Baez:2010ya}
J.~C.~Baez and J.~Huerta,
{\em {An invitation to higher gauge theory},}
\href{http://dx.doi.org/10.1007/s10714-010-1070-9}{Gen. Relativ. Gravit. {\bf
  43}  (2011) 2335} [{\tt \href{http://www.arxiv.org/abs/1003.4485}{1003.4485
  [hep-th]}}].

\bibitem{Zucchini:2015wba}
R.~Zucchini,
{\em {On higher holonomy invariants in higher gauge theory I},}
\href{http://dx.doi.org/10.1142/S0219887816500900}{Int. J. Geom. Meth. Mod.
  Phys. {\bf 13}  (2016) 1650090} [{\tt
  \href{http://www.arxiv.org/abs/1505.02121}{1505.02121 [hep-th]}}].

\bibitem{Zucchini:2015xba}
R.~Zucchini,
{\em {On higher holonomy invariants in higher gauge theory II},}
\href{http://dx.doi.org/10.1142/S0219887816500912}{Int. J. Geom. Meth. Mod.
  Phys. {\bf 13}  (2016) 1650091} [{\tt
  \href{http://www.arxiv.org/abs/1505.02122}{1505.02122 [hep-th]}}].

\bibitem{Caetano:1993zf}
A.~Caetano and R.~F.~Picken,
{\em {An axiomatic definition of holonomy},}
\href{http://dx.doi.org/10.1142/s0129167x94000425}{Int. J. Math. {\bf 5}
  (1994) 835}.

\bibitem{Baez:2004in}
J.~C.~Baez and U.~Schreiber,
{\em Higher gauge theory: 2-connections on 2-bundles,}
{\tt \href{http://www.arxiv.org/abs/hep-th/0412325}{hep-th/0412325}}.

\bibitem{Baez:2005qu}
J.~C.~Baez and U.~Schreiber,
{\em Higher gauge theory,}
\href{http://dx.doi.org/10.1090/conm/431/08264}{Contemp. Math. {\bf 431}
  (2007)~7} [{\tt
  \href{http://www.arxiv.org/abs/math.DG/0511710}{math.DG/0511710}}].

\bibitem{Schreiber:0705.0452}
U.~Schreiber and K.~Waldorf,
{\em Parallel transport and functors,}
J. Homot. Relat. Struct. {\bf 4}  (2009) 187 [{\tt
  \href{http://www.arxiv.org/abs/0705.0452}{0705.0452 [math.DG]}}].

\bibitem{Schreiber:0802.0663}
U.~Schreiber and K.~Waldorf,
{\em Smooth functors vs. differential forms,}
\href{http://dx.doi.org/10.4310/HHA.2011.v13.n1.a7}{Homol. Homot. App. {\bf 13}
   (2011) 143} [{\tt \href{http://www.arxiv.org/abs/0802.0663}{0802.0663
  [math.DG]}}].

\bibitem{Schreiber:2008aa}
U.~Schreiber and K.~Waldorf,
{\em Connections on non-Abelian gerbes and their holonomy,}
\href{http://www.tac.mta.ca/tac/volumes/28/17/28-17.pdf}{Theor. Appl. Categor.
  {\bf 28}  (2013) 476} [{\tt
  \href{http://www.arxiv.org/abs/0808.1923}{0808.1923 [math.DG]}}].

\bibitem{Martins:2007uki}
J.~F.~Martins and R.~Picken,
{\em {On two-dimensional holonomy},}
\href{http://dx.doi.org/10.1090/S0002-9947-2010-04857-3}{Trans. Am. Math. Soc.
  {\bf 362}  (2010) 5657} [{\tt
  \href{http://www.arxiv.org/abs/0710.4310}{0710.4310 [math.DG]}}].

\bibitem{Martins:2011:3309}
J.~F.~Martins and R.~Picken,
{\em Surface holonomy for non-Abelian 2-bundles via double groupoids,}
\href{http://dx.doi.org/10.1016/j.aim.2010.10.017}{Adv. Math. {\bf 226}  (2011)
  3309} [{\tt \href{http://www.arxiv.org/abs/0808.3964}{0808.3964 [math.CT]}}].

\bibitem{Chatterjee:2009ne}
S.~Chatterjee, A.~Lahiri, and A.~N.~Sengupta,
{\em {Parallel transport over path spaces},}
\href{http://dx.doi.org/10.1142/S0129055X10004156}{Rev. Math. Phys. {\bf 22}
  (2010) 1033} [{\tt \href{http://www.arxiv.org/abs/0906.1864}{0906.1864
  [math-ph]}}].

\bibitem{Chatterjee:2014pna}
S.~Chatterjee, A.~Lahiri, and A.~N.~Sengupta,
{\em {Double category related to path space parallel transport and
  representations of Lie 2-groups},}
\href{http://dx.doi.org/10.1007/978-3-642-55361-5_22}{Springer Proc. Math.
  Stat. {\bf 85}  (2014) 379}.

\bibitem{Chatterjee:2010xa}
S.~Chatterjee, A.~Lahiri, and A.~N.~Sengupta,
{\em {Path space forms and surface holonomy},}
\href{http://dx.doi.org/10.1063/1.3275598}{AIP Conf. Proc. {\bf 1191}
  (2009)~66} [{\tt \href{http://www.arxiv.org/abs/1007.1159}{1007.1159
  [math-ph]}}].

\bibitem{Soncini:2014zra}
E.~Soncini and R.~Zucchini,
{\em {A new formulation of higher parallel transport in higher gauge theory},}
\href{http://dx.doi.org/10.1016/j.geomphys.2015.04.010}{J. Geom. Phys. {\bf 95}
   (2015)~28} [{\tt \href{http://www.arxiv.org/abs/1410.0775}{1410.0775
  [hep-th]}}].

\bibitem{Abbaspour:1202.2292}
H.~Abbaspour and F.~Wagemann,
{\em On 2-holonomy,}
{\tt \href{http://www.arxiv.org/abs/1202.2292}{1202.2292 [math.AT]}}.

\bibitem{Abad:1404.0729}
C.~A.~Abad and F.~Schaetz,
{\em Higher holonomies: comparing two constructions,}
\href{http://dx.doi.org/10.1016/j.difgeo.2015.02.003}{Diff. Geo. Appl. {\bf 40}
   (2015)~14} [{\tt \href{http://www.arxiv.org/abs/1404.0729}{1404.0729
  [math.AT]}}].

\bibitem{Abad:1404.0727}
C.~A.~Abad and F.~Schaetz,
{\em Holonomies for connections with values in $L_\infty$-algebras,}
\href{http://dx.doi.org/10.4310/HHA.2014.v16.n1.a6}{Homol. Homot. App. {\bf 16}
   (2014)~89} [{\tt \href{http://www.arxiv.org/abs/1404.0727}{1404.0727
  [math.AT]}}].

\bibitem{Cattaneo:2002tk}
A.~S.~Cattaneo and C.~A.~Rossi,
{\em {Wilson surfaces and higher dimensional knot invariants},}
\href{http://dx.doi.org/10.1007/s00220-005-1339-0}{Commun. Math. Phys. {\bf
  256}  (2005) 513} [{\tt
  \href{http://www.arxiv.org/abs/math-ph/0210037}{math-ph/0210037}}].

\bibitem{Crans:2004ve}
A.~S.~Crans,
{\em Lie 2-algebras,} PhD thesis, University of California Riverside (2004)
[{\tt \href{http://www.arxiv.org/abs/math.QA/0409602}{math.QA/0409602}}].

\bibitem{Crans:1310.4705}
A.~S.~Crans and F.~Wagemann,
{\em Crossed modules of racks,}
\href{http://dx.doi.org/10.4310/HHA.2014.v16.n2.a5}{Homol. Homot. App. {\bf 16}
   (2014)~85} [{\tt \href{http://www.arxiv.org/abs/1310.4705}{1310.4705
  [math.QA]}}].

\bibitem{Zucchini:2011aa}
R.~Zucchini,
{\em {AKSZ models of semistrict higher gauge theory},}
\href{http://dx.doi.org/10.1007/JHEP03(2013)014}{JHEP {\bf 1303}  (2013) 014}
  [{\tt \href{http://www.arxiv.org/abs/1112.2819}{1112.2819 [hep-th]}}].

\bibitem{Zucchini:2015ohw}
R.~Zucchini,
{\em {A Lie based 4-dimensional higher Chern--Simons theory},}
\href{http://dx.doi.org/10.1063/1.4947531}{J. Math. Phys. {\bf 57}  (2016)
  052301} [{\tt \href{http://www.arxiv.org/abs/1512.05977}{1512.05977
  [hep-th]}}].

\bibitem{Soncini:2014ara}
E.~Soncini and R.~Zucchini,
{\em {$4d$ semistrict higher Chern--Simons theory I},}
\href{http://dx.doi.org/10.1007/JHEP10(2014)079}{JHEP {\bf 1410}  (2014)~79}
  [{\tt \href{http://www.arxiv.org/abs/1406.2197}{1406.2197 [hep-th]}}].

\bibitem{hosokawa1979proposals}
F.~Hosokawa and A.~Kawauchi,
{\em Proposals for unknotted surfaces in four-spaces,}
\href{http://dx.doi.org/10.18910/6174}{Osaka J. Math. {\bf 16}  (1979) 233}.

\end{thebibliography}

\bibliographystyle{prop2015}

\end{document}